\documentclass[twocolumn,floats,amsmath,amssymb,nofootinbib]{aastex7}
\usepackage{bm}
\usepackage[english]{babel}
\usepackage{tabu}
\usepackage{amsmath}
\usepackage[caption=false, position=t, singlelinecheck=off]{subfig}
\usepackage[utf8]{inputenc} 
\usepackage[T1]{fontenc} 
\usepackage{multirow}
\usepackage[encapsulated]{CJK}
\usepackage{ucs}

\DeclareRobustCommand{\cntext}[1]{\begin{CJK}{UTF8}{bkai}#1\ignorespacesafterend\end{CJK}} 

\received{}
\revised{}
\accepted{}
\submitjournal{ApJ}

\shorttitle{Gravitational Collapse of Small Dust Grains}
\shortauthors{Baehr et al. 2025}

\graphicspath{{./}{figures/}}

\makeatletter
\newcommand*\bigcdot{\mathpalette\bigcdot@{.5}}
\newcommand*\bigcdot@[2]{\mathbin{\vcenter{\hbox{\scalebox{#2}{$\m@th#1\bullet$}}}}}
\makeatother

\begin{document}

\title{On the Gravitational Collapse of Small Dust Grains in Self-gravitating Disk Structures}

\citestyle{egu}
\bibliographystyle{yahapj}

\correspondingauthor{Hans Baehr}
\email{baehr@mpia.de}

\author[0000-0002-0880-8296]{Hans Baehr}
\affil{Department of Physics and Astronomy, The University of Georgia, Athens, GA 30602, USA}
\affil{Center for Simulational Physics, The University of Georgia, Athens, GA 30602, USA}
\affil{Max Planck Institute for Astronomy, Königstuhl 17, D-69117 Heidelberg, Germany}
\email{baehr@mpia.de}

\author[0000-0002-6379-9185]{Ken Rice}
\affil{SUPA, Institute for Astronomy, University of Edinburgh, The Royal Observatory, Blackford Hill, Edinburgh, EH93HJ, UK}
\affil{Centre for Exoplanet Science, University of Edinburgh, Edinburgh, EH9 3HJ, UK}
\email{wkmr@roe.ac.uk}

\author[0000-0003-2589-5034]{Chao-Chin Yang (\cntext{楊朝欽})}
\affil{Department of Physics and Astronomy, The University of Alabama, Box~870324, Tuscaloosa, AL~35487-0324, USA}
\email{ccyang@ua.edu}

\author[0000-0003-2589-5034]{Cassandra Hall}
\affil{Department of Physics and Astronomy, The University of Georgia, Athens, GA 30602, USA}
\affil{Center for Simulational Physics, The University of Georgia, Athens, GA 30602, USA}
\email{cassandra.hall@uga.edu}

\begin{abstract}
Planet formation may begin much earlier than previously expected, when the protoplanetary disk is still massive and gravitationally unstable. It has been proposed that solid grains can concentrate in the spiral arms of self-gravitating disks, leading to the formation of planetary embryos or cores that can greatly accelerate the process of planet formation. We perform hydrodynamic simulations of self-gravitating gas and even smaller dust grains than previously investigated in 3-dimensional shearing box simulations to explore the conditions necessary to form these planetary seeds. Focusing on small grains of dimensionless stopping time $\mathrm{St}=0.01$ and shorter, we find that disk metallicities $Z \gtrsim 0.02$ can overcome the disruptive effects of dust diffusion among these small dust grains. In the outer reaches of a gravitationally unstable disk, these models correspond to grains of approximately 1$\,mm$ and lead to planetary embryos between 0.1 and 1 Earth mass. The formation of these planetary embryos could therefore reduce the time needed for planet assembly, particularly in the outer regions of the disk where coagulation timescales are longer and solid growth is limited.
\end{abstract}


\section{Introduction}
\label{sec:intro}

Circumstellar disks less than one million years old may be sites of ongoing planet formation, as indicated by the presence of disk structures in disks approximately one to three million years old \citep{Andrews2018a}. This is further supported by the growing evidence that planets may exist within some of the gaps such as those inferred through gas kinematics \citep{Pinte2018,Teague2018}. These potential planet detections estimate planetary masses around a few Jupiter masses typically at a few tens of au, which would indicate that planet formation needs to occur rapidly, even at wide orbital separations.

Rapid planet formation at wide orbits has been the domain of gravitational instabilities (GI) and disk fragmentation, whereby a massive enough disk can be unstable to self-gravity of the gas into spiral structures \citep{Shu1987} or bound fragments \citep{Boss1997}. Fragmentation can concentrate large amounts of gas on time scales comparable to the orbital timescale, making it a viable alternate mechanism to form gas giants \citep{Boss1997}. Not only does there need to be a significant amount of gas, but it needs to cool rapidly so that pressure support cannot prevent gravitational collapse; otherwise the disk goes into a gravitoturbulent state without fragmentation \citep{Gammie2001}. Thus, gravitational fragmentation of the disk has been used to explain the formation of gas giant planets beyond $\sim$ 30 au \citep{Rafikov2005,Janson2012,Vigan2017}.

However, disk fragmentation likely leads to companions with masses greater than 10 $M_{Jup}$ \citep{Kratter2016}, making it unlikely that GI can explain objects suspected of inhabiting the gaps of protoplanetary disks \citep{Zhang2018}. An alternative exists when one considers the possibility of the dust in the disk reaching the point of gravitational instability. The gravitational collapse of dust directly into planetesimals was originally considered in \citet{Goldreich1973}, who found that Kelvin-Helmholtz instabilities prevented the collapse of thin dust layers, unless the disk metallicity were sufficiently increased \citep{Youdin2002}. Numerical studies have since considered the dust of predominantly larger grain sizes, either assuming that planetesimals have already formed by the time gravitational instabilities develop \citep{Boley2010} or dust grains are on the order of centimeters or larger \citep{Rice2004,Gibbons2012,Gibbons2015}.

In recent years it has become apparent that for these larger grain sizes, gravitational collapse of the dust can result in bodies larger than planetesimals, up to several Earth masses in size \citep{Baehr2022,Longarini2023b,Rowther2024}. However, dust grains larger than a few millimeters in radius are probably not present in young disks where gravitational instabilities are most likely, particularly at wide orbital separations \citep{Ohashi2022}. Compared to larger grain sizes, which can quickly and efficiently concentrate within spiral arms \citep{Gibbons2014a,Shi2016}, smaller grains pose a few challenges to this process. First, small grains settle to the disk midplane slowly, and form a more extended, less concentrated vertical layer. Second, small grains drift towards the pressure maximum at the center of the spiral density perturbation much slower than larger particles that are not as well coupled to the gas. At canonical interstellar medium (ISM) metallicity $Z = M_{d}/M_{g} \sim 0.01$, this implies that small dust may remain diffuse, reducing the chance of reaching the densities necessary for gravitational collapse. However, at higher dust loads even small dust grains should be able to concentrate to levels that lead to the formation of planetary precursors. Forming a planetary core at an early stage of the disk lifetime could shorten the formation time of a detectable planet at tens of au to within 1 Myr \citep{Baehr2023}.

With this study, we explore the behavior of small grains with hydrodynamic shearing box simulations of a self-gravitating disk. In particular, we focus on what enhancements to the disk metallicity can produce bound concentrations of dust that may serve as seeds for planet formation. The paper continues with an overview of the theory of dusty self-gravitating disks in Section \ref{sec:theory} followed by the details of the hydrodynamical simulations in Section \ref{sec:model}. In Section \ref{sec:results} we present the results and discuss the implications on planet formation and other topics in Section \ref{sec:discussion}.

\section{Theory}
\label{sec:theory}

Self-gravitating disks should have enough mass such that even the thermal pressure and rotation of the disk cannot prevent the gas from collapsing. This is captured in the so-called Toomre parameter \citep{Safronov1960,Toomre1964,Goldreich1965}
\begin{equation}
Q \approx \frac{c_{\mathrm{s}}\Omega}{\pi G \Sigma},
\end{equation}
where $G$ is the gravitational constant, $c_{s}$ is the gas sound speed, $\Omega$ is the Keplerian frequency and $\Sigma$ is the vertically-integrated gas density.

As suggested in our previous study \citep{Baehr2022}, dust particles in a gravitoturbulent disk can collapse under their own gravity if the diffusive motions of the particles are weak enough compared to the self-gravity of the local concentration. A range of intermediate particle sizes were explored at canonical metallicity to evaluate a particle collapse criterion that could be applied to the diffusive particle motions introduced by the streaming instability \citep{Gerbig2020}. In Equation \eqref{eq:gerbigparameter} below, we define a dust stability criterion $Q_{\mathrm{d}}$ which is a modification of the gas stability criterion $Q$ considering that dust particles within the gas are collisionless and unaffected by the gas pressure \citep{Gerbig2020}. In this case, the internal diffusion of the dust $\sqrt{\delta_x/\mathrm{St}}$ resulting from coupling to the gas turbulence acts as a pressure-like term for the dust component. We use the radial dimensionless diffusion constant $\delta_x$ as a substitute for $\delta$ as it is typically an order of magnitude larger and dominates the diffusion. This constant is normalized by the disk scale height $H$ and sound speed, $\delta_x = D_x/Hc_s$ where the diffusion in the radial direction $D_x$ can be evaluated by
\begin{equation} \label{eq:diffusionconstant}
D_x \equiv \frac{1}{2}\frac{d\langle | x(t) - x(0) |^{2} \rangle}{dt}\,.
\end{equation}
The particle size is represented here by the dimensionless Stokes number $\mathrm{St} = \tau_{\mathrm{s}}\Omega$, which characterizes the coupling time between a dust particle and the gas in terms of the stopping time $\tau_{\mathrm{s}}$. When a concentration of dust at a distance $R$ around a star of mass $M_*$ is enhanced by $\epsilon = \Sigma_{\mathrm{d}}/\langle \Sigma_{\mathrm{d}} \rangle \sim 100$, the dust mass approaches densities comparable to the Hill density,
\begin{equation} \label{eq:hilldensity}
\rho_{\mathrm{Hill}} = \frac{9}{4\pi}\frac{M_*}{R^3}.
\end{equation}
Thus, the particle layer can become gravitationally unstable\footnote{Because we consider the radial diffusion $\delta_x$ we omit the 3/2 factor which is necessary when considering isotropic collapse.} with a collapse criterion
\begin{equation} \label{eq:gerbigparameter}
Q_{\mathrm{d}} = \frac{Q}{\epsilon Z}\sqrt{\frac{\delta_x}{\mathrm{St}}} < 1.
\end{equation}
This is slightly different from the model of \citet{Goldreich1973} in that it considers the diffusion strength of the particles as a result of gas turbulence \citep{Klahr2020,Klahr2021}. Although Equation \eqref{eq:gerbigparameter} was shown to adequately explain the formation of bound clouds of dust particles between $\mathrm{St} = 0.1$ to $\mathrm{St} = 10$, young, gravitationally unstable disks are unlikely to have significant quantities of dust grains in these larger sizes \citep{Booth2016,Ohashi2023a,Han2023,Aso2025}. Therefore the motivation of this work is to explore the dust concentration of the smaller size particles that are more likely to exist in young disks.

In the context of the streaming instability and the formation of planetesimals it has been shown before that metallicity has a strong impact on the formation of bound objects, with similar constraints on the necessary metallicity \citep{Johansen2009,Yang2017,Li2021}. However, the strength of particle diffusion in disks subject to marginal gas gravitational instabilities is different to that of streaming instabilities, often by a few orders of magnitude \citep{Shi2016}. In an attempt to understand which dust sizes are most likely to collapse and under what conditions in a gravitationally turbulent disk, we recast Equation \eqref{eq:gerbigparameter} as the level of diffusion that allows for the collapse of dust
\begin{equation} \label{eq:diffusionlimit}
\delta \lesssim \mathrm{St}(\epsilon Z)^2.
\end{equation}
We have assumed that the gas stability is of order $Q \approx 1$. If the enhancement in the surface density of the particle disk required for collapse is around $\epsilon = 100$, as suggested by \citet{Shi2016} and \citet{Rice2025}, then the simple relation $\delta < \mathrm{St}$ means that as particle size decreases, weaker levels of diffusion can be enough to prevent the dust from collapsing into a bound cloud. This also implies however, that as dust size increases, there is no amount of diffusion that can prevent gravitational collapse and clumping should be seen for particle sizes $\mathrm{St}\gtrsim 1$, which is not the case \citep{Baehr2022}. We suggest large dust with $\mathrm{St} \gtrsim 1$ cannot efficiently concentrate in dense spiral density waves due to the weak coupling to the gas and the stronger gravitational interactions between large dust and the spirals, which dominates over drag terms, leading to a gravitational stirring effect \citep{Shi2016}. 

The relationship between gravitational effects and gas drag is given by \citep{Shi2016,Baehr2021a}
\begin{equation} \label{eq:stirringlimit}
\frac{a_{\mathrm{drag}}}{a_{\mathrm{grav}}} \approx \frac{Q}{\mathrm{St}},
\end{equation}
where $a_{\mathrm{drag}}$ and $a_{\mathrm{grav}}$ are the accelerations due to drag and the gravitational force of a spiral respectively. Combined with a delayed response to the dynamic, transient gas density perturbations inherent with GI ($Q \sim 1$), this relation suggests that dust sizes larger than $\mathrm{St} \gtrsim 1$ are less likely to move towards a pressure maximum and more likely to take a trajectory deflected by the gravity of a spiral perturbation.

\section{Model}
\label{sec:model}

\begin{deluxetable*}{ccccccccccccc}
\tablecaption{List of Simulations:\label{tab:sims}}
\tablehead{
\colhead{Model} & \colhead{$\mathrm{St}$} & \colhead{$\beta$} & \colhead{$Z$} & \colhead{$Q$} & \colhead{$\alpha_R$} & \colhead{$\alpha_G$} & \colhead{$H_{d}\, [H_{g}]$} & \colhead{$\delta_{d,x}$}& \colhead{$\delta_{d,z}$} & \colhead{ $\sigma_{d,x}$ $[c_s]$} & \colhead{ $\sigma_{d,z}$ $[c_s]$ }}
\startdata
small\_Z001 & $10^{-2}$ & $10$ & $0.01$ & $1.3$ & $1.0\times 10^{-2}$ & $1.1\times 10^{-3}$ & $0.12$ & $1.0\times 10^{-2}$ & $9.6\times 10^{-4}$ & $0.25$ & $0.11$\\
small\_Z002 & $10^{-2}$ & $10$ & $0.02$ & $1.3$ & $7.7\times 10^{-3}$ & $1.5\times 10^{-3}$ & $0.11$ & $7.2\times 10^{-3}$ & $4.5\times 10^{-5}$ & $0.24$ & $0.10$\\
small\_Z004 & $10^{-2}$ & $10$ & $0.04$ & $1.3$ & $1.3\times 10^{-2}$ & $1.4\times 10^{-4}$ & $0.17$ & $5.0\times 10^{-3}$ & $8.6\times 10^{-5}$ & $0.27$ & $0.10$\\
xsmall\_Z004 & $10^{-3}$ & $10$ & $0.04$ & $1.3$ & $1.2\times 10^{-2}$ & $7.8\times 10^{-4}$ & $0.15$  & $7.4\times 10^{-3}$ & $3.4\times 10^{-2}$ & $0.26$ & $0.10$ \\
\enddata
\tablecomments{Simulation parameters $\mathrm{St}$, cooling timescale $\beta$ and metallicity $Z$ and steady-state values of the stability parameter $Q$, Reynolds stress $\alpha_R$, gravitational stress $\alpha_G$, dust scale height $H_{d}$ relative to the gas scale height $H_{g}$, dimensionless particle diffusion constants $\delta$ and particle velocity dispersions $\sigma$, the values of which are averaged over $t = 30\Omega^{-1}$ to $t = 100\Omega^{-1}$. All simulations have spatial resolution $512^{2}\times 256$ with box lengths $L_{x}=L_{y}=(80/\pi) H_g$ and $L_{z}=(40/\pi) H_g$. Simulations are initially marginally gravitationally stable such that the gas $Q_{0} = 1.02$ and then evolve to the indicated self-regulating value of $Q$. For measured quantities within the gravitationally particle clouds, see Table \ref{tab:clumps}.}
\end{deluxetable*}

We use the same simulation setup as those presented in \citet{Baehr2022}, using the PENCIL\footnote{http://pencil-code.nordita.org/}  code \citep{Brandenburg2003,PencilCode2021} to model local shearing boxes of self-gravitating dust and gas. However, we now explore smaller grains ($\mathrm{St} < 0.1$) which require small timesteps at high solid concentrations. To model the dust-gas interaction of small dust particles without being too severely hampered by the numerical timestep, we use the method introduced in \citet{Yang2016}. In short, this method uses a operator-splitting scheme to separate the shear, rotation and dust feedback terms of the gas momentum equation into a new differential equation that is integrated locally on a cell-by-cell basis as a system along with the particle velocity equation. An analytical solution can be obtained in this case and hence the method relieves the time-step constraints imposed by the stopping time and local dense concentrations of the particles. As a result, modeling the motion of dense concentrations of small particles becomes more tractable.

Gas and dust particles both contribute to and are affected by the gravitational potential, and the backreaction force of particles onto the gas is calculated by mapping the change of particle momentum due to the gas drag back to the grid with triangular-shaped clouds \citep{Youdin2007}. All simulations are resolved with $512^{2}\times 256$ grid cells and box lengths $L_{x}=L_{y}=(80/\pi) H_g$ and $L_{z}=(40/\pi) H_g$ such that $\Delta x=\Delta y =\Delta z \simeq 0.05\, H_g$, where $H_g$ is the vertical scale height of the gas at the initial uniform temperature. Simulations are run using 512 processors up to a simulation time of $t = 100\,\Omega^{-1}$.

In a shearing box, the hydrodynamic equations of the co-rotating disk system are linearized and transformed into Cartesian coordinates, where $q = -d\mathrm{ln}\Omega / d\mathrm{ln}R = 3/2$ is the shear parameter:
\begin{align}
\frac{\partial {\rho_{\mathrm{g}}}}{\partial t} &- q\Omega x\frac{\partial {\rho_{\mathrm{g}}}}{\partial y} + \nabla\cdot(\rho_{\mathrm{g}}\bm{u}) = f_{D}(\rho_{\mathrm{g}}) \label{eq:finalmassconserve} \\
\frac{\partial \bm{u}}{\partial t} &- q\Omega x\frac{\partial \bm{u}}{\partial y} + \bm{u}\cdot\nabla\bm{u} = -\frac{\nabla p}{\rho_{\mathrm{g}}} + q\Omega u_{x}\bm{\hat{y}} \nonumber \\ 
& - 2\bm{\Omega}\times\bm{u} - \nabla\Phi - \mathbf{g} - \frac{\varepsilon}{\tau_{s}} ( \bm{u} - \bm{w} ) + f_{\nu}(\bm{u}) \label{eq:finalmomconserve} \\
\frac{\partial s}{\partial t} &-q\Omega x\frac{\partial s}{\partial y} + (\bm{u} \cdot \nabla)s = -\frac{\Lambda}{\rho_{\mathrm{g}} T} + f_{\chi}(s). \label{eq:finalenergyconserve}
\end{align}
In the above, $\mathbf{u}$ is the deviation of the gas velocity from the background shear velocity, $\bm{w}$ is the particle velocity which imparts a backreaction onto the gas proportional to the local dust-to-gas ratio $\varepsilon \equiv \rho_\mathrm{d} / \rho_\mathrm{g}$, $\rho_{\mathrm{g}}$ is the gas density, and $\rho_{\mathrm{d}}$ is the dust density. The energy equation evolves the specific entropy $s$, while $p$ is the gas pressure and $T$ is the gas temperature. We use a vertical gravitational acceleration $\bm{g} = g\hat{\bm{z}}$ which is a linear profile modified with zero acceleration near the $z$-boundary to avoid an abrupt discontinuity at the periodic vertical boundary. All simulations use a cooling timescale $t_c = \beta \Omega^{-1}$ such that fragmentation of the gas will not occur. We refer to \citet{Baehr2022} for the details of the hyperviscosities ($f_D$, $f_\nu$ and $f_\chi$) and cooling function ($\Lambda$), which are unchanged here.

Every superparticle in the simulation represents a collection of solids with identical properties such that the $i$-th superparticle has position $\bm{x}^{(i)}$ and velocity $\bm{w}^{(i)}$ as in \citet{Youdin2007,Yang2016}. The equations that govern the particle position and velocity are
\begin{align}
\frac{d \bm{w}^{(i)}}{d t} &= -2\Omega \times \bm{w}^{(i)} +q\Omega w^{(i)}_x\bm{\hat{y}} - \nabla\Phi \nonumber\\
&- \mathbf{g} - \frac{1}{\tau_{s}} \left( \bm{w}^{(i)}  - \bm{u}(\bm{x}^{(i)}) \right), \label{eq:particlevel}\\
\frac{d \bm{x}^{(i)}}{d t} &= \bm{w}^{(i)} - q\Omega x^{(i)} \bm{\hat{y}}. \label{eq:particlepos}
\end{align}
If the particles are smaller than the mean free path of the gas, the drag they feel is in the Epstein regime, and the stopping time is proportional to the particle size as \citep{Weidenschilling1977}
\begin{equation} \label{eq:epsteinregime}
\tau_{\mathrm{s}} = \frac{a\rho_{\bigcdot}}{c_{\mathrm{s}}\rho_{\mathrm{g}}},
\end{equation}
where $a$ is the radius of the particle and $\rho_{\bigcdot}$ is the material density of an individual dust particle.

Particles are initially placed within the simulation with a Gaussian vertical profile with the same width as that of the gas but random in the $x$-$y$ plane. For our base model, the cumulative dust mass is equally distributed among all particles such that the total mass is one, two or four percent that of the gas, corresponding to a metallicity of $Z=0.01, 0.02, 0.04$ respectively.

We continue using the same scaling relations and diagnostics as in \citet{Baehr2022}, which assumes that the shearing box is centered at a radial position of $R= 50\,au$.  We repeat the scalings relations below. 
\begin{equation}
c_s = 199 \left( \frac{T}{11.25\,K} \right)^{1/2} \, \mathrm{m\, s}^{-1}
\end{equation}
\begin{equation}
\Omega = 1.78 \times 10^{-2} \left( \frac{R}{50\, \mathrm{au}} \right)^{-3/2} \left( \frac{M_{*}}{1 M_{\odot}} \right)^{1/2}\, yr^{-1}
\end{equation}
\begin{equation}
P = 353 \left( \frac{R}{50\, \mathrm{au}} \right)^{3/2} \left( \frac{M_{*}}{1 M_{\odot}} \right)^{-1/2}\, yr
\end{equation}
\begin{equation}
H_{g} = 2.36 \left( \frac{T}{11.25\,K} \right)^{1/2} \left( \frac{R}{50\, \mathrm{au}} \right)^{3/2} \left( \frac{M_{*}}{1 M_{\odot}} \right)^{-1/2}\, \mathrm{au}.
\end{equation}
This gives a unit mass of
\begin{equation} \label{eq:codemassunit}
\begin{split}
\hat{M}_{0} &= \frac{4}{\pi}\frac{H_{g}^{3}}{GP^{2}}\\
&=6.8 \times 10^{27} \left( \frac{T}{11.25\,K} \right)^{3/2} \left( \frac{R}{50\, \mathrm{au}} \right)^{3/2}\\ 
&\times\left( \frac{M_{*}}{1 M_{\odot}} \right)^{-1/2}\, \mathrm{g} = 1.1\,M_{\oplus}.
\end{split}
\end{equation}
With this value for code unit of mass, we derive the total gas mass in the entire simulation box (assuming the values from above at 50 au)
\begin{equation}
\begin{split}
M_{\mathrm{total,gas}} &= \Sigma_{0} L_{x}L_{y} = 80^{2} \hat{M}_{0}/Q_0 \\
&= 4.3 \times 10^{31}\, g = 0.021\,M_{\odot}.
\end{split}
\end{equation}
The total dust mass varies according to the metallicity $Z$ as shown in Table \ref{tab:sims}, so that
\begin{equation}
\begin{split}
M_{\mathrm{total,dust}} &= 80^2 Z \hat{M}_{0}/Q_0 \\
&= 4.3 \times 10^{29}\,\left( \frac{Z}{0.01} \right) \mathrm{g} \\
&= 72\,\left( \frac{Z}{0.01} \right)M_{\oplus}.
\end{split}
\end{equation}
This mass is distributed equally among all super particles ($N_p = 1.5 \times 10^{6}$) yielding a mass per super particle
\begin{equation}
\begin{split}
M_{\mathrm{sp}} &= 2.8 \times 10^{23}\,\left( \frac{Z}{0.01} \right) \mathrm{g} \\ 
&= 4.8 \times 10^{-5}\,\left( \frac{Z}{0.01} \right) M_{\oplus}.
\end{split}
\end{equation}
The above scaling relations allow us to convert the code units to physical units at any desired radius of the disk, which for our analysis is at $R = 50$~au. In the following section, we look at the results of our simulations, analyzing the mass of clumps that form as well as under what conditions of metallicity and particle size they can form.

\section{Results}
\label{sec:results}

\begin{figure*}[t]
\centering
\includegraphics[width=0.5\textwidth]{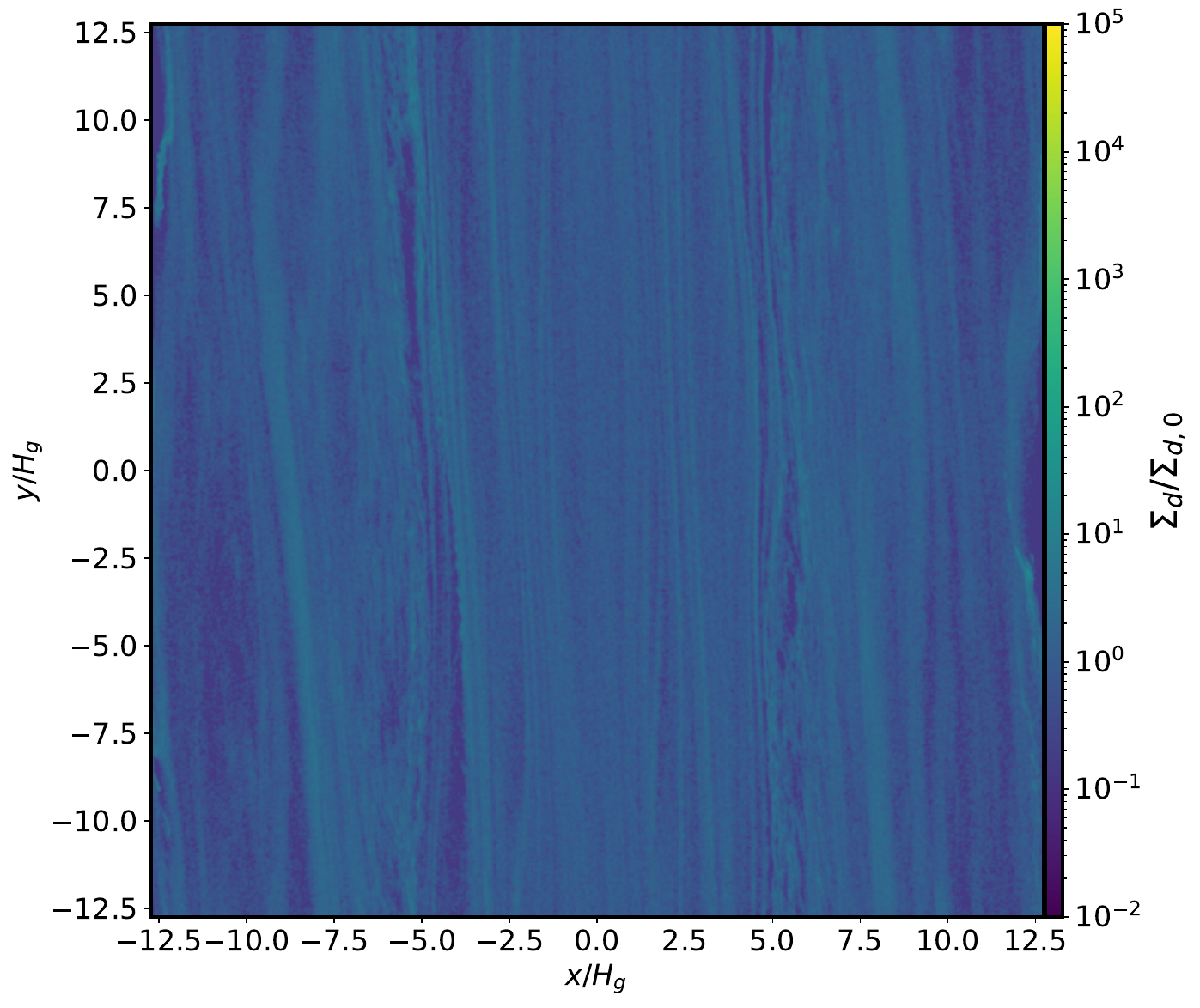}%
\includegraphics[width=0.5\textwidth]{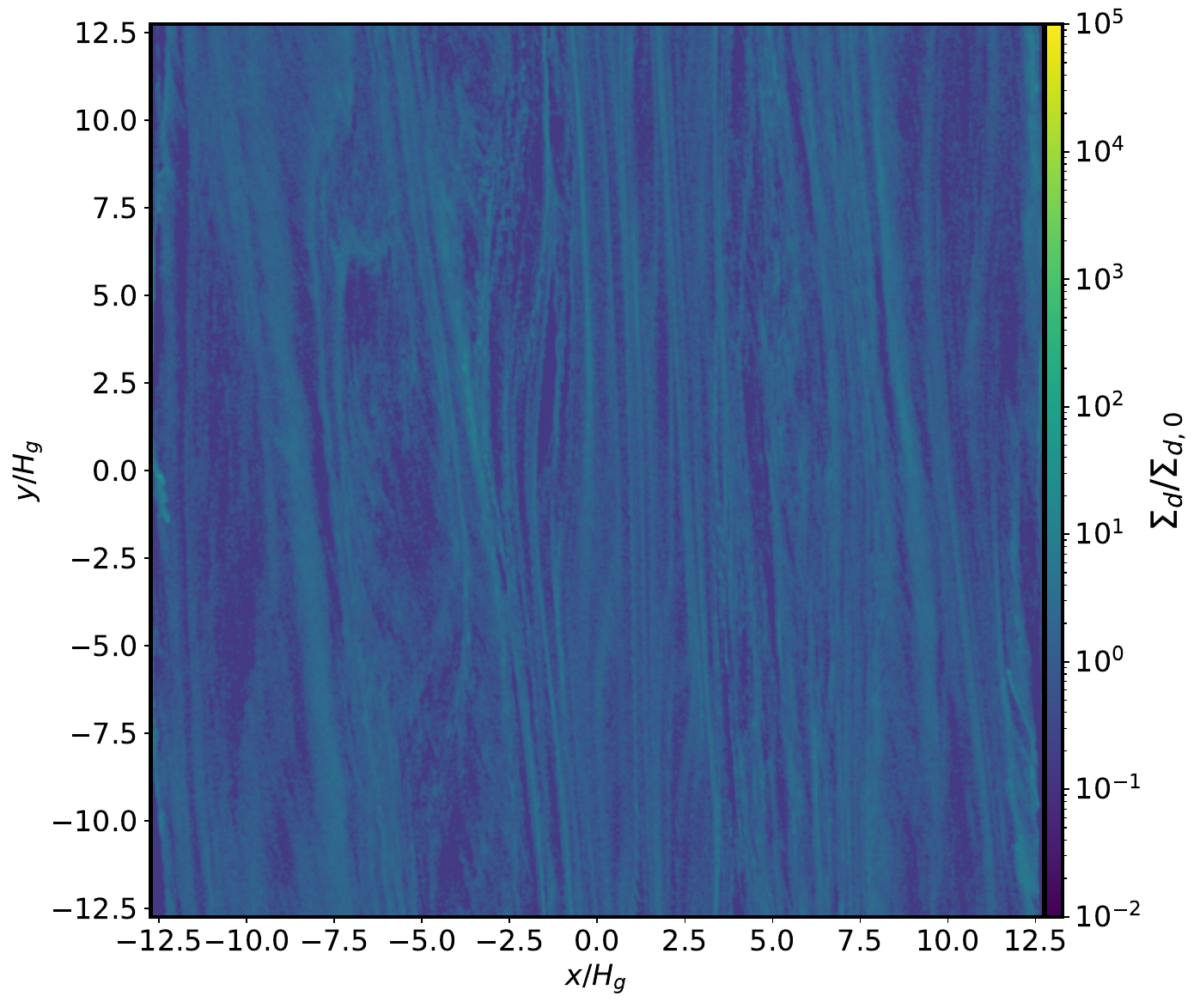}
\includegraphics[width=0.5\textwidth]{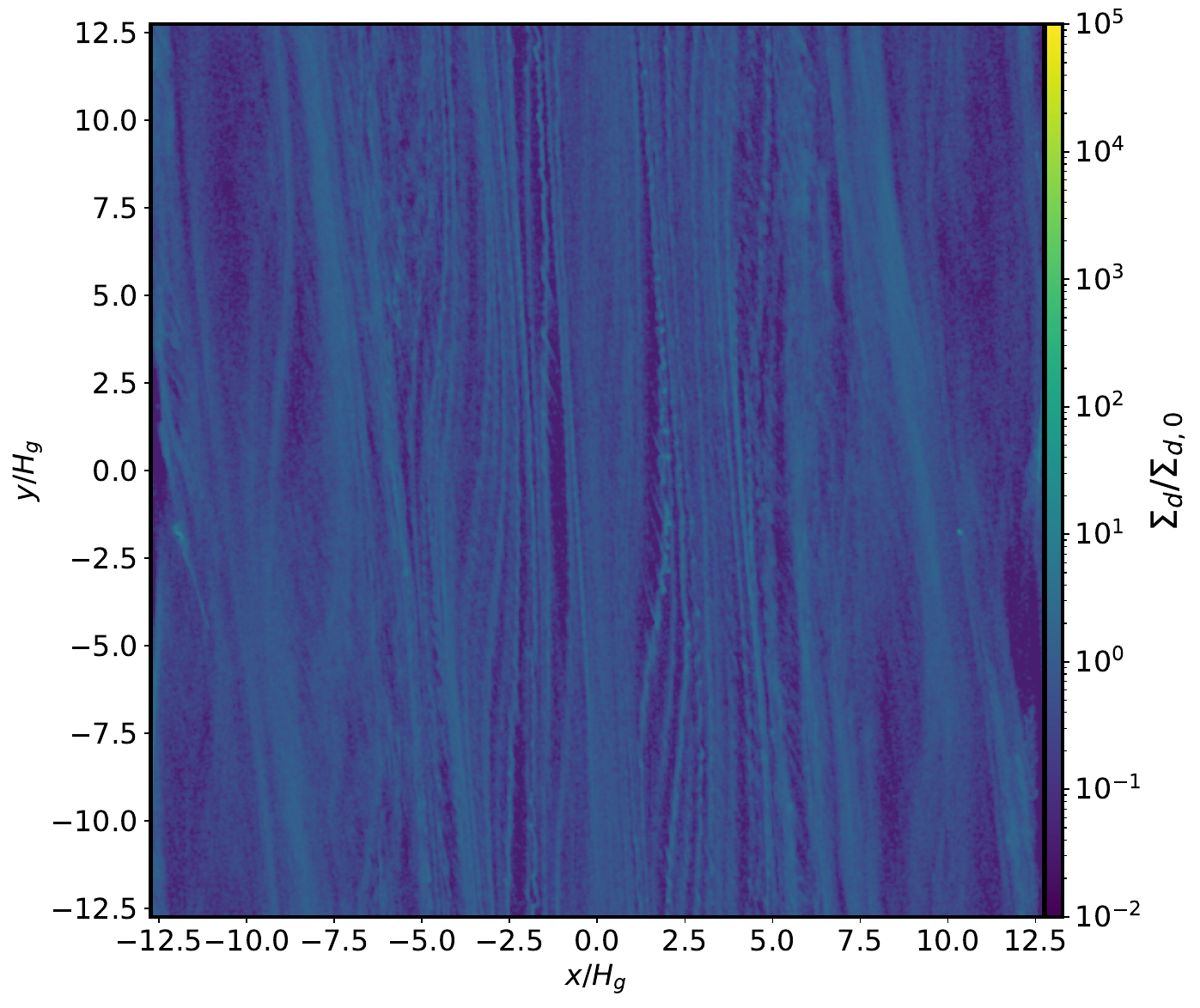}%
\includegraphics[width=0.5\textwidth]{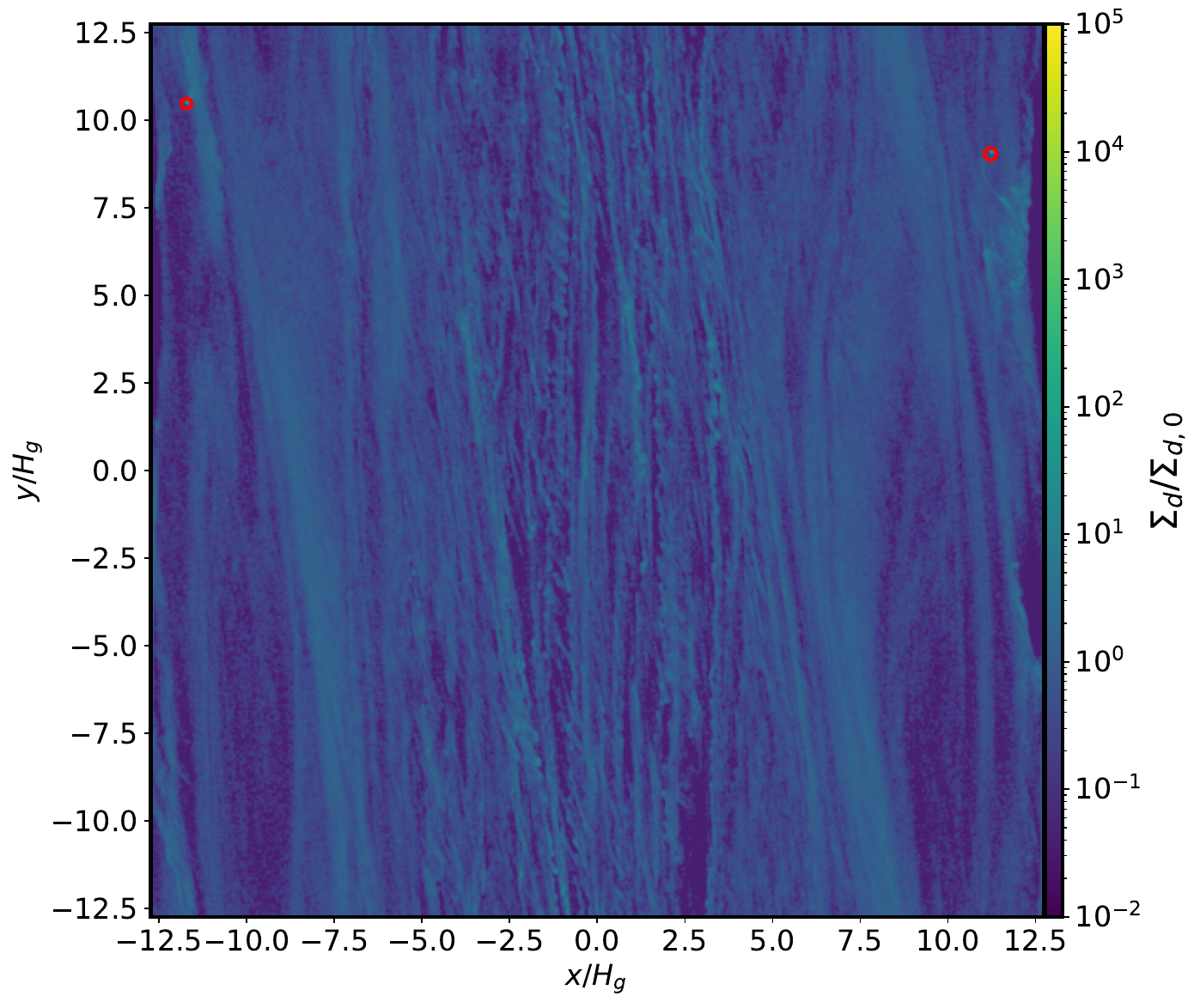}
\includegraphics[width=0.5\textwidth]{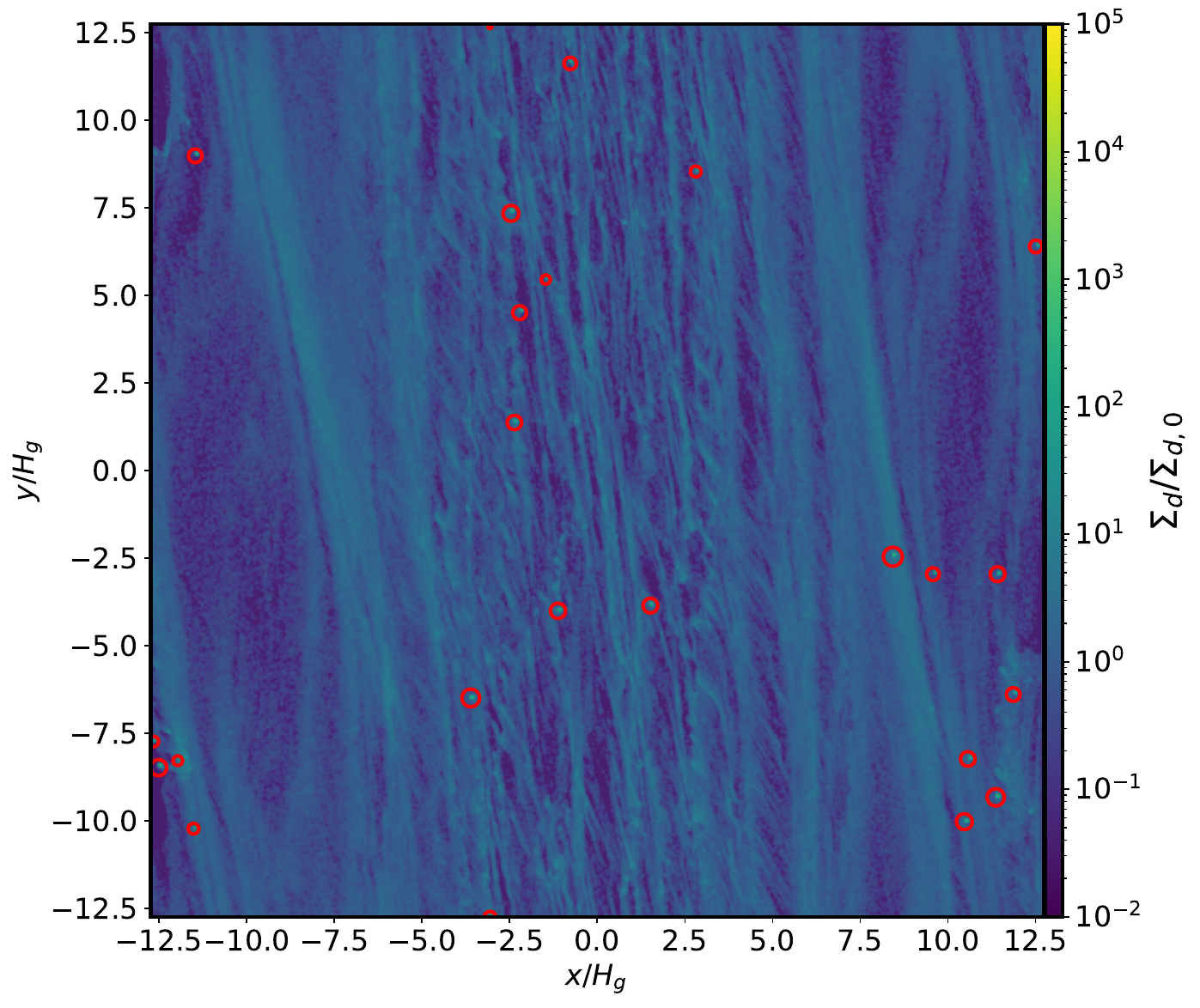}%
\includegraphics[width=0.5\textwidth]{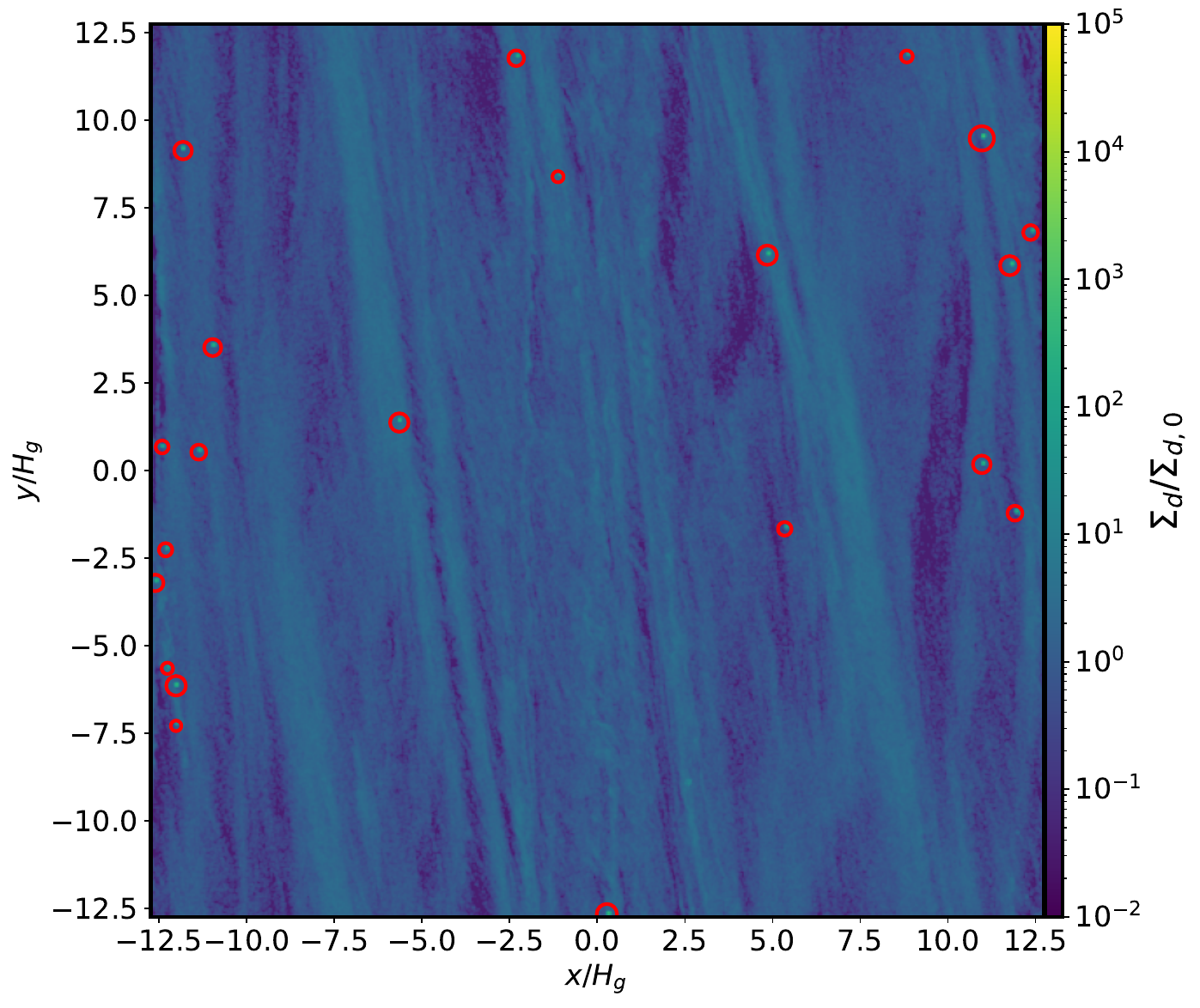}
\caption{Snapshots of the dust surface density enhancement in three simulations with different total metallicities, at two different times: $t = 65\,\Omega^{-1}$ (left column) and $t = 90\,\Omega^{-1}$ (right column). The simulation shown in the top row is the case where $Z=0.01$ and the dust does not gravitationally collapse within the simulation runtime. The simulation shown in the middle row uses dust with metallicity $Z=0.02$ while the bottom row has $Z=0.04$. Dense clouds of dust above Roche surface density (Eq. \eqref{eq:rochedensity}) are indicated with red circles.}
\label{fig:ST001maps}
\end{figure*}

\begin{figure*}[t]
\centering
\includegraphics[width=0.5\textwidth]{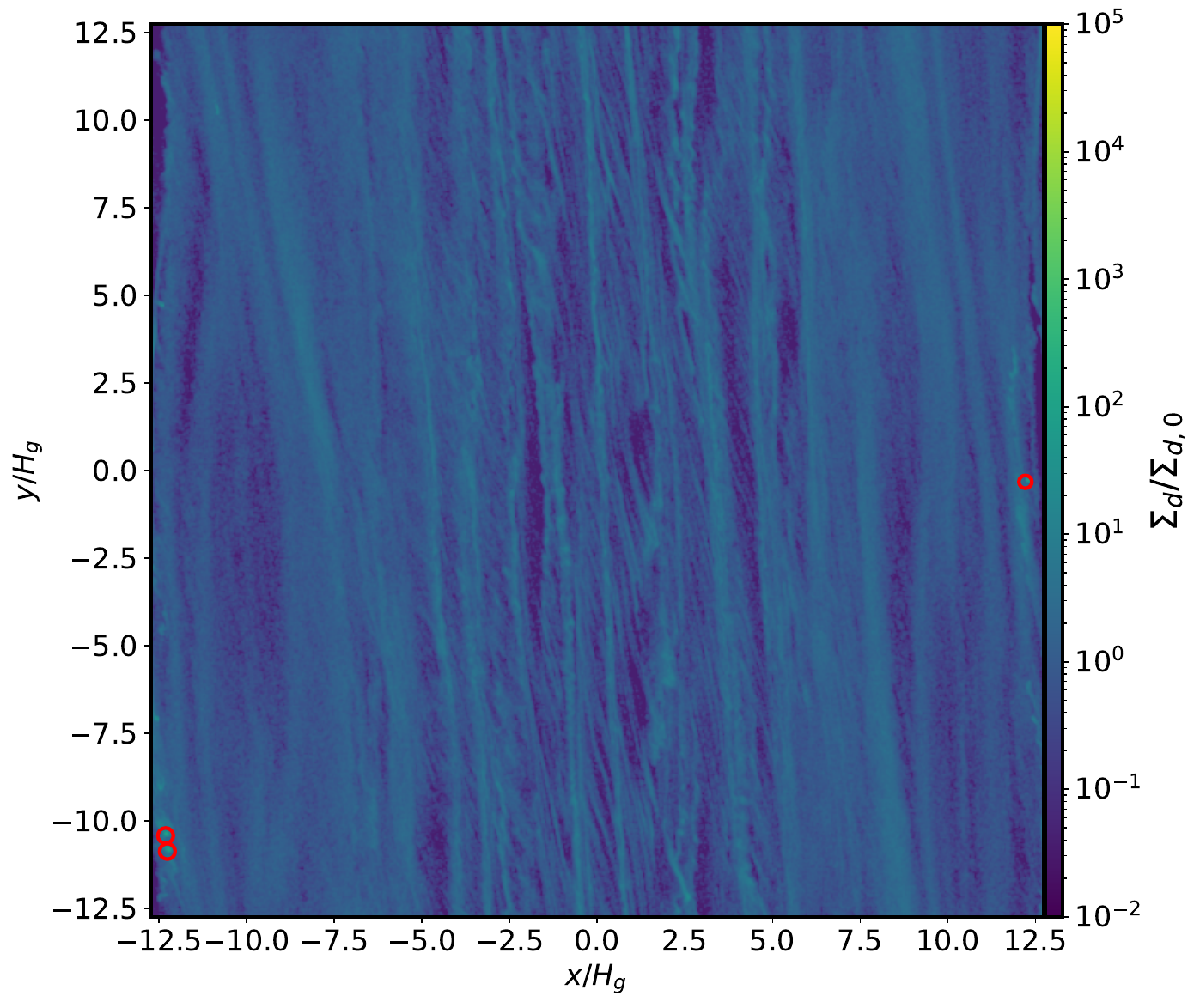}%
\includegraphics[width=0.5\textwidth]{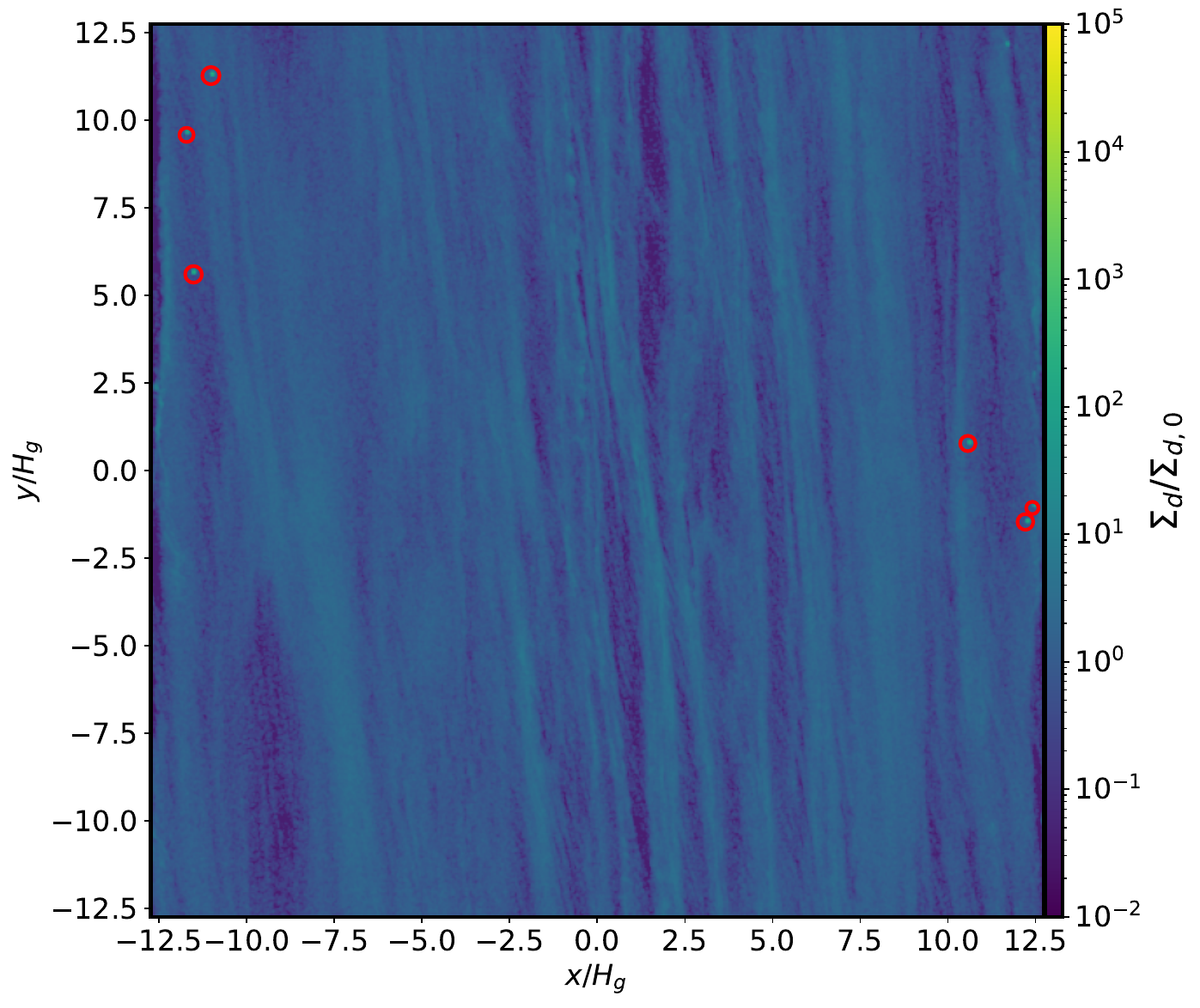}
\caption{Same as Figure \ref{fig:ST001maps}, but for the simulation with $\mathrm{St} = 0.001$ and $Z=0.04$.}
\label{fig:ST0001maps}
\end{figure*}

\begin{figure}[t]
\centering
\includegraphics[width=0.47\textwidth]{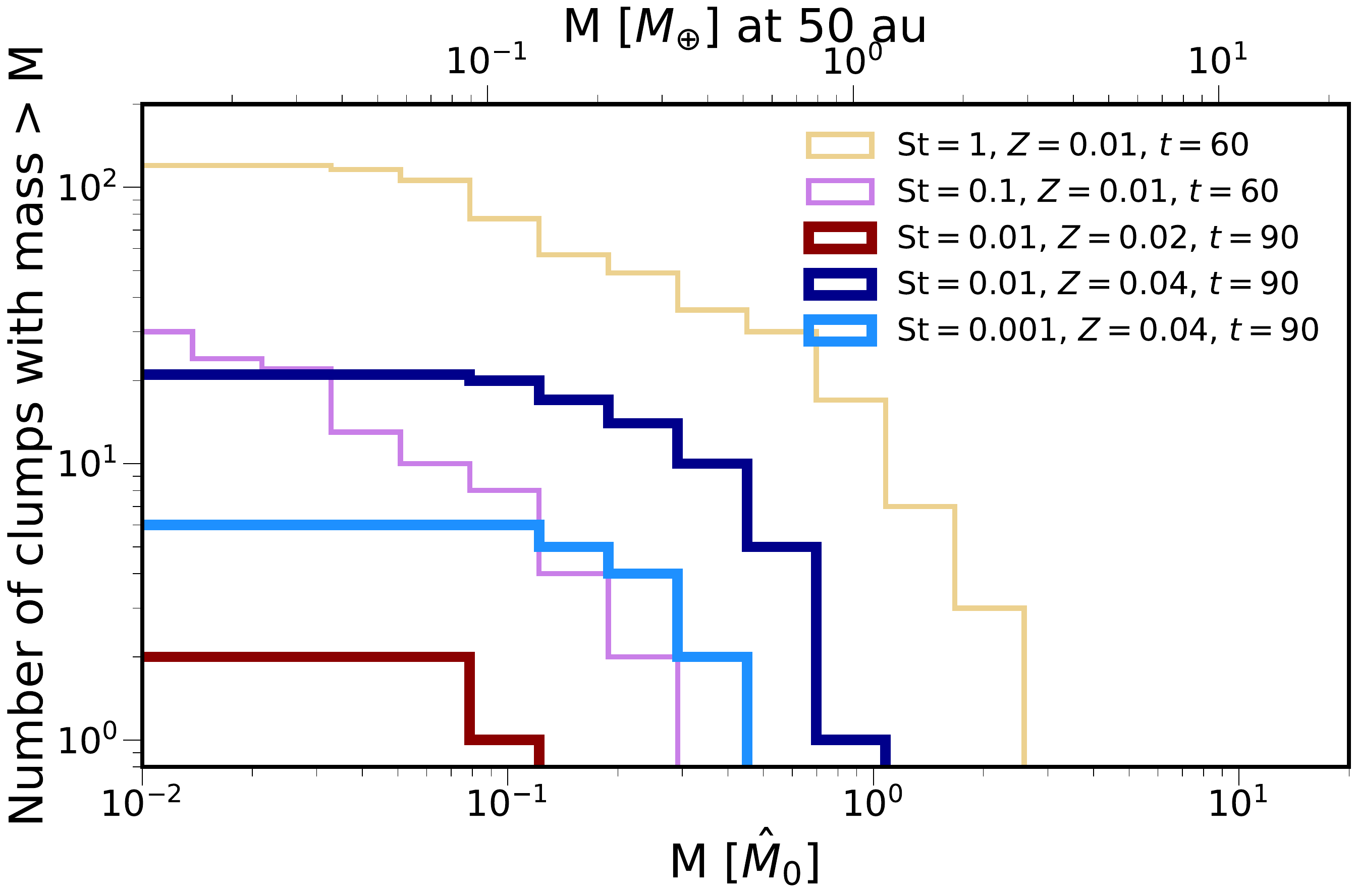}
\caption{The cumulative distribution of bound dust clumps formed in the simulations at $t=90\Omega^{-1}$. The flat part of the distributions on the left-hand side of the plot indicates that there are no clumps forming below this mass. Compared to the simulations with larger dust sizes in which clumps can form below $\hat{M}$ (thinner, paler lines) \citep[e.g.,][]{Baehr2022}, smaller grains require stronger local dust concentrations to overcome the dust diffusion.}
\label{fig:massdistribution}
\end{figure}

In Figure \ref{fig:ST001maps} we show the enhancement of the vertically integrated particle density in our simulations with $\mathrm{St} = 0.01$ particles for three different initial metallicities at two different times. The top row shows metallicity $Z=0.01$, the nominal value in the interstellar medium. As anticipated from Equation \eqref{eq:gerbigparameter} \citep{Baehr2023}, dense dust clumps are not able to form as the dust is unable to reach the necessary local density enhancement to overcome the diffusion of the dust. The middle and bottom rows show the results with metallicities $Z=0.02$ and $Z=0.04$, respectively. When the metallicity is increased to $Z=0.02$, there is more mass in the dust component leading to higher concentrations within the dense gas structures, resulting in the formation of a pair of dust clumps after several orbits. Only once the metallicity is increased by another factor of two to $Z=0.04$ are clumps formed promptly after the initial collapse phase and in greater quantities. Figure \ref{fig:ST0001maps} shows the same snapshots for the simulation with $\mathrm{St} = 0.001$ particles with metallicity $Z=0.04$, which forms a handful of bound clumps. Clumps are identified and indicated with red circles once they maintain a Roche surface density threshold as derived in \citet{Baehr2022}
\begin{equation} \label{eq:rochedensity}
\Sigma_{R} \approx 8.8\frac{\Omega^{2}H_g}{G}.
\end{equation}

This indicates that there is a regime where dense clouds of small dust grains can form and potentially seed planet formation, even if dust growth is negligible or large grains are mostly converted into smaller grains by collisions \citep{Booth2016}. This trend extends to yet smaller sizes $\mathrm{St} = 10^{-3}$, although the number of clumps is lower and the onset of collapse is later compared to $\mathrm{St} = 0.01$ grains.

In Figure \ref{fig:massdistribution}, we plot the mass distribution of clumps formed in our simulations, assuming the shearing box is centered at a radius of 50 au. For comparison, the distribution of the larger particles from \citet{Baehr2022} are plotted in fainter, thinner lines. In the simulation with the highest metallicity, most clumps are in the range of 0.1 to 1 $M_{\oplus}$. The minimum clump mass for simulations with $\mathrm{St}=0.01$ grains is the same regardless of the metallicity. The minimum mass of a clump for $\mathrm{St}=0.001$ grains is even larger by a factor of approximately 2 (see Table \ref{tab:clumps}), suggesting that smaller grains need to reach higher density enhancements to form dense clouds. This is expected if the contained mass is regulated by the diffusion of the dust particles and more dust has to concentrate for smaller grains to collapse. This is a result of the timescale of collapse increasing as particle sizes decrease such that diffusion has more time to prevent the formation of a bound cloud \citep{Klahr2020}. This suggests that there is some limiting grain size where the amount of mass needed for collapse to proceed is larger than is locally available and formation of bound clumps may no longer reliably occur.

\begin{figure*}[t]
\centering
\includegraphics[width=0.47\textwidth]{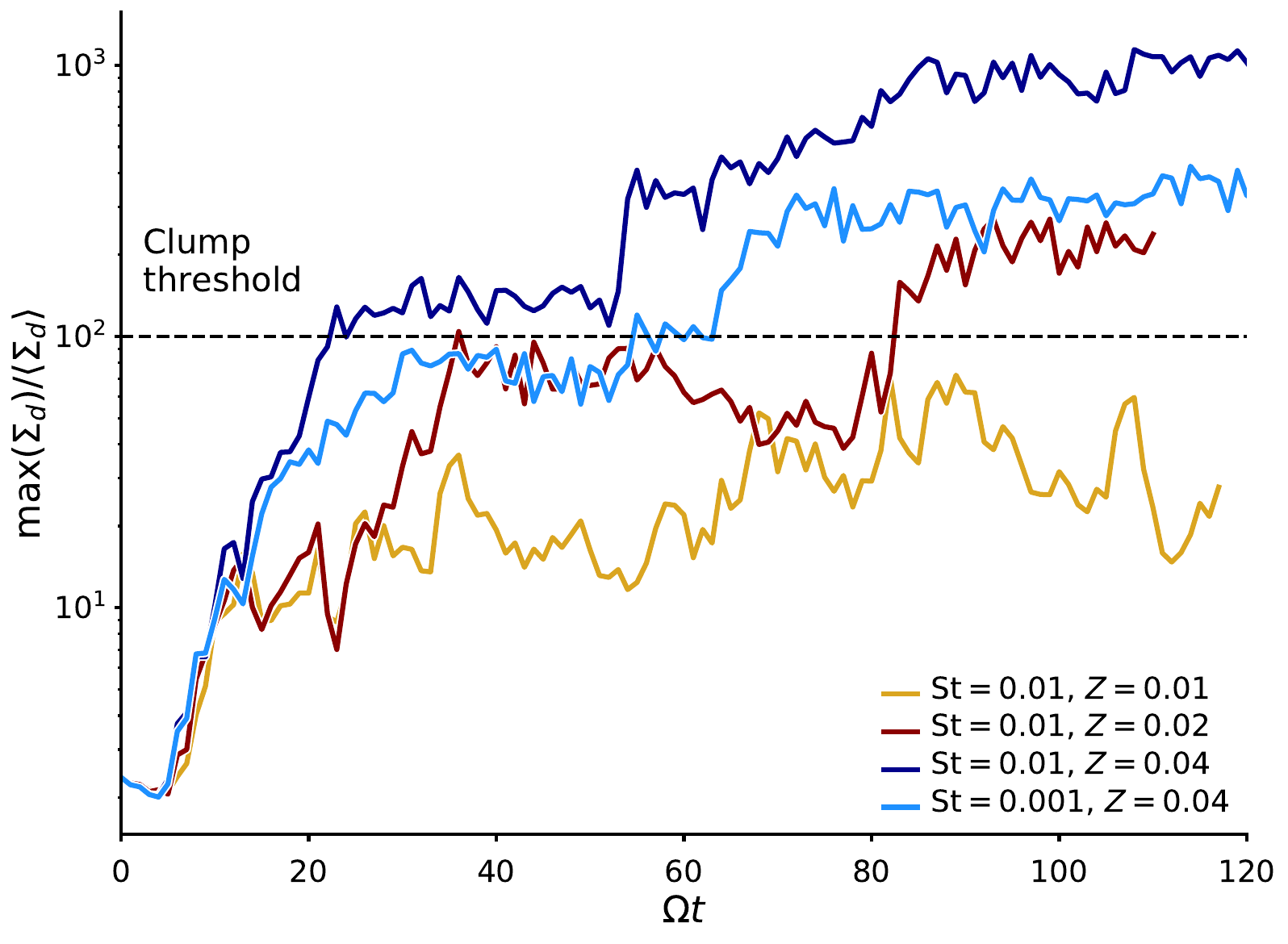}%
\includegraphics[width=0.5\textwidth]{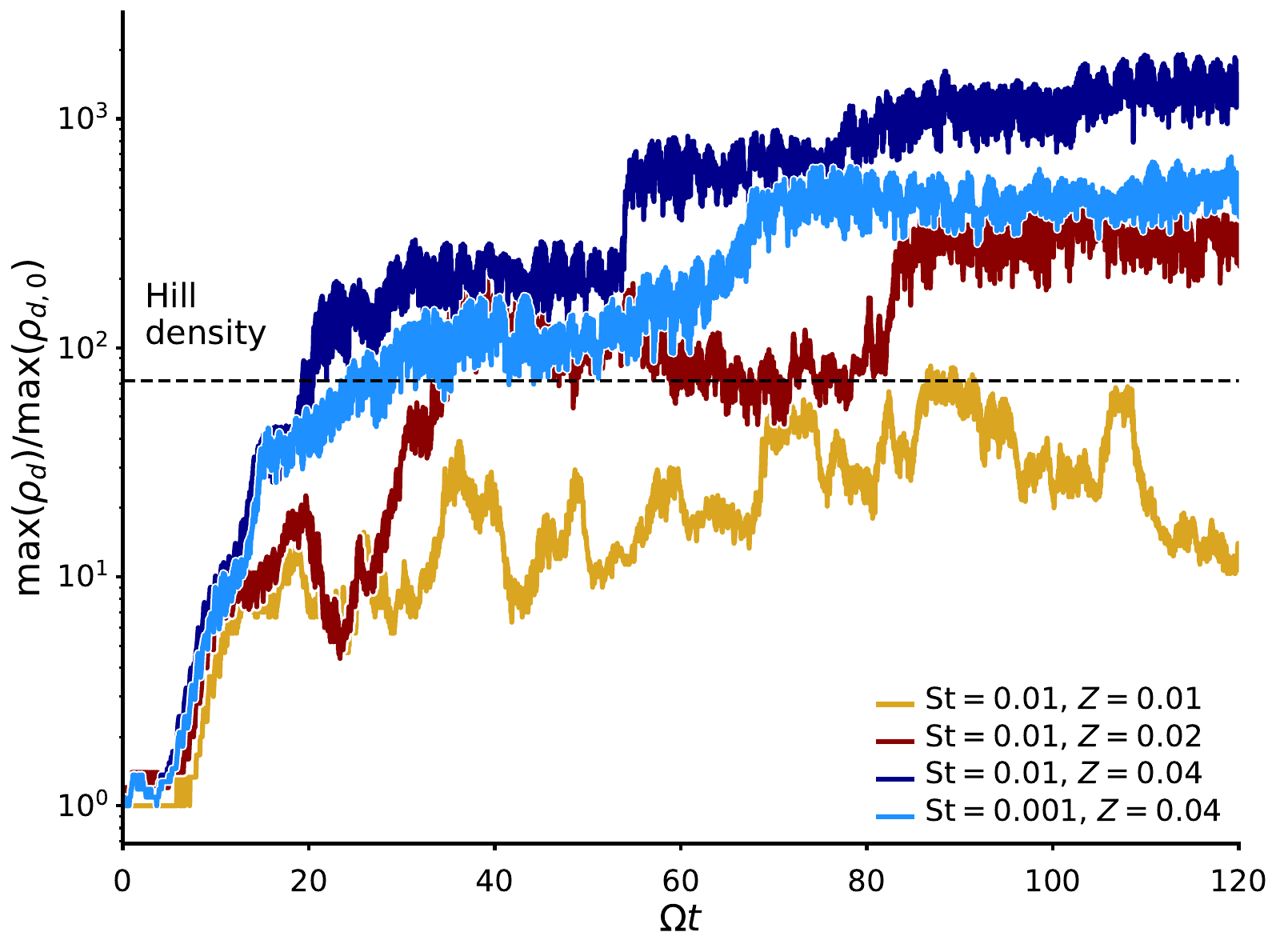}
\caption{\emph{Left:} The time evolution of the maximum dust surface density $\Sigma_{d}$ relative to the spatial average $\langle \Sigma_{d} \rangle$. The approximate threshold separating the surface densities which result in dense clouds is marked with a horizontal dashed line at $\Sigma_d = 100$. \emph{Right:} The time evolution of the maximum dust density $\rho_{d}$ relative to the initial density maximum $\rho_{d,0}$. The Hill density is indicated by the horizontal dashed line.}
\label{fig:densityevolution}
\end{figure*}

Dust clouds that form are limited to around 1 $\hat{M}_0$ and when more clumps form they also form up to a larger mass. If dust clouds are easy to form and are numerous they can merge with nearby clouds soon after formation, thus increasing the number of higher mass objects. However, in these simulations, clumps tend to be fairly isolated and interactions or mergers are rare. Additionally, within a gas filament concentrations are higher (for example, for particles closer to $\mathrm{St}=1$) and it is easier to surpass the necessary local dust enhancement and create more dense clouds above a particular size. Since small grains concentrate and settle on longer timescales, collapse proceeds once the minimum mass needed for collapse is reached, although some additional growth may occur while the clump forms.

In the left panel of Figure \ref{fig:densityevolution}, we compare the evolution of the maximum dust surface density enhancement over time in each simulation. In the case where we have particles of $\mathrm{St} = 0.01$ and assume ISM metallicity $Z=0.01$, the surface density initially increases by a factor of 10 to 20 before later rising to around 30 to 40. Over the course of the simulation this is not enough to trigger gravitational collapse of the dust. Increasing the metallicity by a factor of two similarly initially reaches enhancements over 40 in $\Sigma_d$, but still has difficulty reaching a density where it can collapse into a bound cloud. However, after $t = 80 \Omega^{-1}$ a pair of small bound dust clumps eventually form but do not grow or change significantly during the remainder of the run. Increasing the dust mass further to $Z=0.04$ sees the simulation reach 100 times the average dust surface density by $t = 20\Omega^{-1}$ which forms a handful of initial clumps before forming more after $t = 50\Omega^{-1}$. Increases in the dust surface density when $Z=0.04$ and $\mathrm{St} = 10^{-3}$ are not as pronounced when compared to the larger particle size with the same metallicity but are still higher than the lower metallicities.

The right panel of Figure \ref{fig:densityevolution} shows the maximum of the dust volume density as a function of time compared with the Hill density. While the simulations where the maximum density stay above the Hill density eventually form bound dust clumps, it does not mean that it has occurred. For example, between times $t = 30\Omega^{-1}$ and $t = 60\Omega^{-1}$, three curves are above the Hill density, but only the dark blue ($\mathrm{St} = 0.01$ and $Z=0.04$) simulation forms bound clumps of dust during this interval, indicating the effects of dust diffusion.

\begin{figure}[t]
\centering
\includegraphics[width=0.46\textwidth]{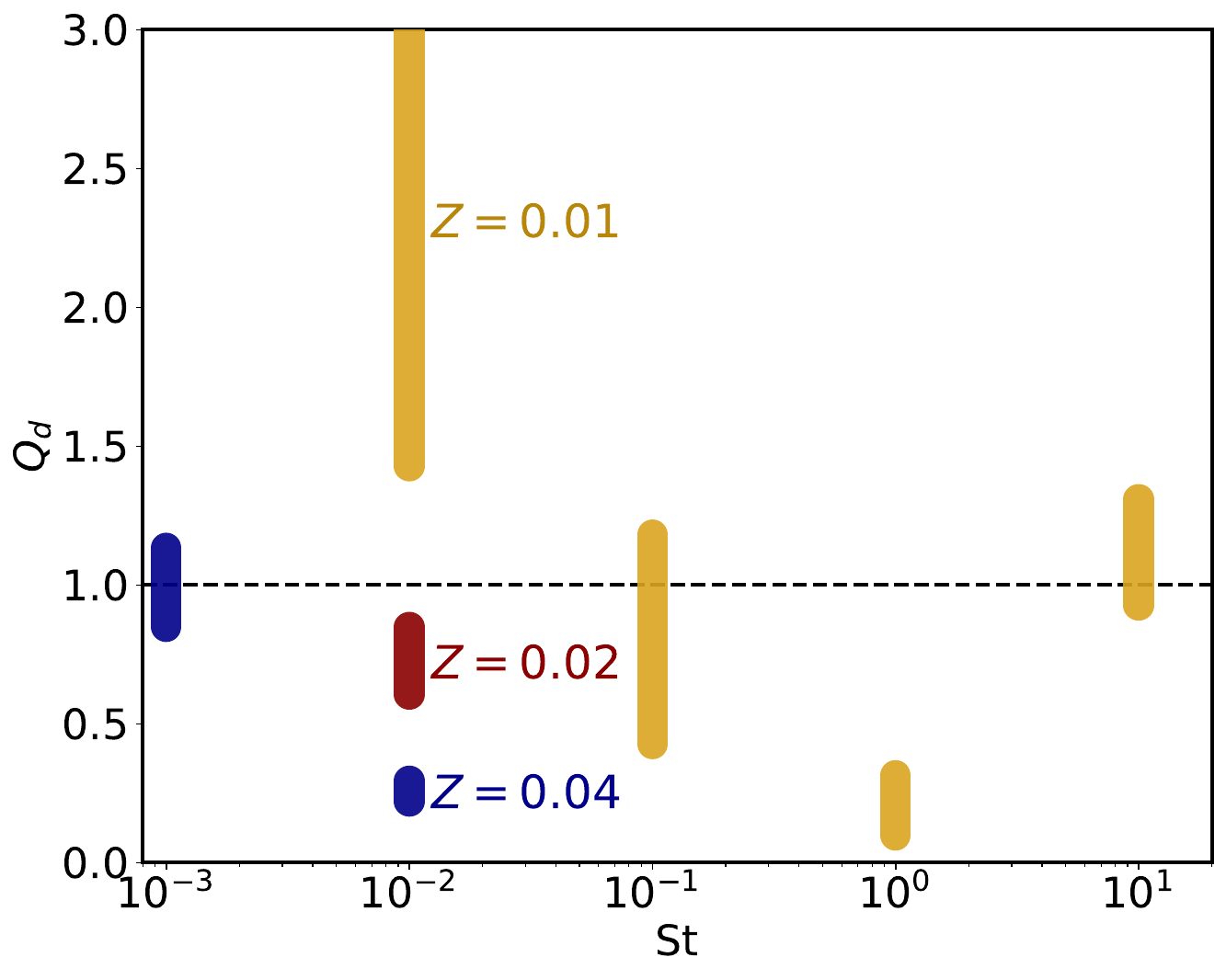}
\caption{The value of $Q_d$ based on measured values of $\delta$ and $\epsilon$ for various particle sizes. The values for $\mathrm{St} \geq 0.1$ come from \citet{Baehr2022} and those for $\mathrm{St} = 0.01$ and $\mathrm{St} = 0.001$ are newly measured in this work. Increasing the metallicity for $\mathrm{St} = 0.01$ leads to values of $Q_d$ consistent with the gravitational collapse observed. In the case of $\mathrm{St} = 0.001$, the value of $Q_d$ is in the range of stability, even though the dust does gravitationally collapse.}
\label{fig:particleQ}
\end{figure}

\begin{deluxetable*}{cccccccccccc}
\tablecaption{Particle clump properties at $t = 90\Omega^{-1}$:\label{tab:clumps}}
\tablehead{
\colhead{model} & \colhead{$N$} & \colhead{ \begin{tabular}[c]{@{}l@{}} $\langle M_{\mathrm{dust}}\rangle$  $[\hat{M}_{0}]$ \end{tabular}} & \colhead{ \begin{tabular}[c]{@{}l@{}} $\langle M_{\mathrm{gas}}\rangle$ $[\hat{M}_{0}]$ \end{tabular}} & \colhead{$\langle Z \rangle$} & \colhead{\begin{tabular}[c]{@{}l@{}} $\langle\sigma\rangle$ $[c_s]$ \end{tabular}} & \colhead{\begin{tabular}[c]{@{}l@{}} $M_{\mathrm{min}}$ $[\hat{M}_{0}]$ \end{tabular}}& \colhead{ \begin{tabular}[c]{@{}l@{}} $M_{\mathrm{max}}$ $[\hat{M}_{0}]$ \end{tabular}} & \colhead{$Q_d$} }
\startdata
small\_Z001 & $0$ & -- & -- & -- & -- & -- & -- & 2.6 \\
small\_Z002 & $2$ & $0.09$ & $0.09$ & $1.0$ & $6.2\times 10^{-2}$ & $6.5\times 10^{-2}$ & $0.12$ & 0.71 \\
small\_Z004 & $21$ & $0.30$ & $0.12$ & $2.5$ & $0.16$ & $6.9\times 10^{-2}$ & $0.94$ & 0.25 \\
xsmall\_Z004 & $6$ & $0.23$ & $0.09$ & $2.5$ & $0.12$ & $0.12$ & $0.33$ & 1.0 \\
\enddata
\tablecomments{Summary of clump properties and particle properties within bound clumps in each simulation at $t=90\Omega^{-1}$, where $N$ is the number of identified bound objects, $M_{\mathrm{dust}}$ and $M_{\mathrm{gas}}$ are the dust and gas mass within the Hill sphere of each clump, $Z = M_{\mathrm{dust}}/M_{\mathrm{gas}}$,  $\sigma$ is the particle velocity dispersion. Quantities in $\langle\cdot\rangle$ brackets are averaged over all clumps in the simulation. The stability criterion $Q_d$ is measured in the time leading up to clump formation for most simulations, from $t = 20 \Omega^{-1}$ to $t = 50 \Omega^{-1}$. The simulation with $\mathrm{St}=0.01$ and $Z=0.01$ `small\_Z001' did not form any bound clumps, but is included for completeness. Masses are in units of $\hat{M}_{0}=(4/\pi) G^{-1}H^{3}_{g}P^{-2}$ (Equation \eqref{eq:codemassunit}).} 
\end{deluxetable*}

\section{Discussion}
\label{sec:discussion}

As discussed above, Figure \ref{fig:densityevolution}, shows the evolution in time of the maximum enhancement in dust surface density $\Sigma_d/\langle \Sigma_{d} \rangle$, which is the enhancement factor $\epsilon$ defined in Equation \eqref{eq:gerbigparameter}. Figure \ref{fig:densityevolution} suggests that $\Sigma_d/\langle \Sigma_d \rangle \sim 100$ separates the times when the simulations have formed dense clouds and the times when they have not. If the gas is close to gravitational instability so that $Q \approx 1$ and the dust-to-gas ratio is $0.01$, then a roughly 100-fold enhancement in the dust surface density will lead to gravitational instability of the dust layer. This, however, does not appear to be supported by these simulations. One might expect simulations with higher metallicities should require a weaker dust concentration enhancement in order to reach Hill density and form clumps, and they instead need the same enhancement. We find no evidence of higher levels of turbulence at higher metallicities that might affect the ability of dust to collapse. The conditions under which collapse is easiest, $\mathrm{St} = 0.01$ and $Z=0.04$, produce a number of early clumps which only accumulate more particles later. This is a result of the high initial turbulence of the initial gas collapse phase which produces additional diffusion in the dust particles. Once this subsides and diffusion decreases, the clouds of dust collapse further. In contrast, similar simulations by \citet{Riols2020} found no evidence of particle clumping, using a fluid approximation for the dust compared the Lagrangian particles employed here. From these simulations they measured thicker dust layers and higher diffusion for $\mathrm{St} < 1$ compared to \citet{Baehr2021a}, which might affect clumping at small $\mathrm{St}$.

When it comes to dust concentration, grain sizes smaller than $\mathrm{St} = 1$ may present a problem because the thickness of the dust layer becomes greater as the grain size decreases \citep{Yang2018}. While $\mathrm{St}=1$ grains settle quickly and form a thin, dense layer, small grains with near perfect coupling to the gas have a thickness comparable to that of the gas \citep{Youdin2007a}. This is partially alleviated by the self-gravity of the disk, but is largely due to the gas and not the particles themselves \citep{Baehr2021}. We find that increasing the metallicity has a negligible impact on the dust scale height of the small particles considered in this study and if anything the dust layer is very slightly thicker with increased metallicity (see Table \ref{tab:sims}).

Figure \ref{fig:particleQ} shows the measured value of $Q_d$ in each simulation compared to the simulations from \citet{Baehr2022}. The strength of diffusion among dust particles sets the initial size of objects formed through gravitational collapse \citep{Klahr2020}. In the case of the streaming instability, dust concentrates gradually, such that upon reaching the critical value of $Q_d$, collapse occurs promptly into bodies of similar sizes \citep{Gerbig2020,Gerbig2023}. For a self-gravitating disk, where concentration due to drift and self-gravity can occur more rapidly, it is possible that more dust concentrates before gravitational collapse occurs and values of $Q_d < 1$ are reached, resulting in a range of initial masses.

We find that clumps form with similar average masses $\hat{M} \approx 0.1 - 0.3$ but with higher minimum sizes and lower maximum sizes than those at $\mathrm{St} \geq 0.1$ from corresponding simulations in \citet{Baehr2022}. The result is a much narrower distribution of bodies in smaller quantities. This can be explained by the additional dust mass required to overcome dust diffusion, which is what keeps the dust layer fairly thick. When compared to simulations of $\mathrm{St} = 0.1$ particles from \citet{Baehr2022} there is however little impact on the maximum size of the clumps. The total number of clumps decreases as the total amount of mass required to instigate collapse becomes more difficult to reach. For instance, as the metallicity is decreased for particle size $\mathrm{St}=0.01$ we see that $Q_d$ increases and the number of clumps within the simulation decreases. This is consistent with the idea that a certain mass needs to be reached within a particular volume, and the lower the value of $Q_d$ the easier it is to reach or exceed this threshold. Furthermore, all simulations here include dust feedback, which was found in \citet{Baehr2022} to make more smaller clumps, perhaps indicating that feedback can make clumping easier, although the impact on the clump mass distribution was not dramatic.

\begin{figure*}[t]
\centering
\includegraphics[width=0.70\textwidth]{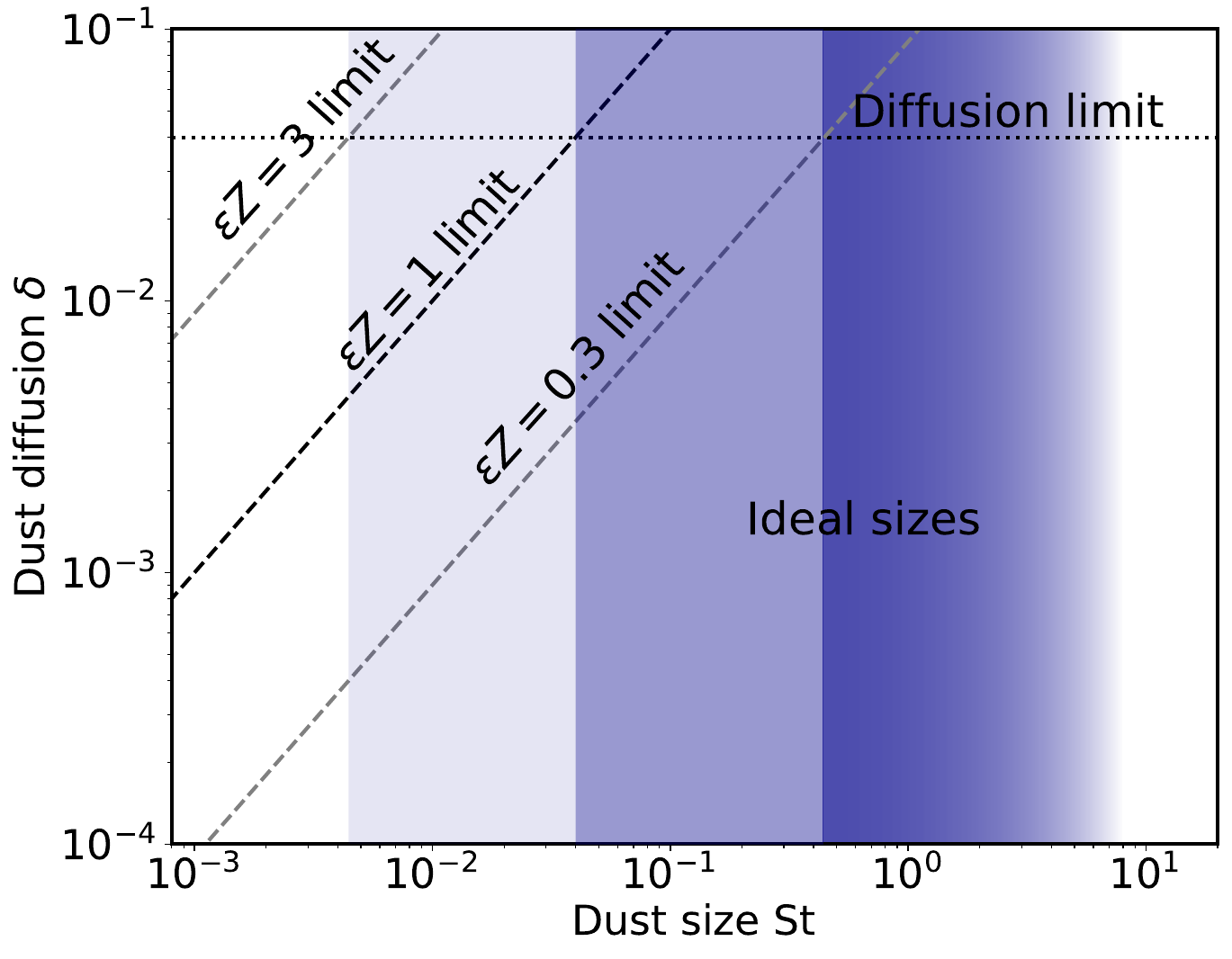}
\caption{A sketch of what dust sizes might be expected to gravitationally collapse for differing levels of diffusivity and with different assumptions of the disk metallicity. The limitation at the higher dust sizes is the regime where coupling is weak and gravitational interactions between dust and the spiral arms are more consequential than drift towards the pressure maximum. Using Equation \eqref{eq:diffusionlimit} and assuming $\varepsilon = 100$, we create diagonal dashed lines which for three different metallicities limit collapse to the region below each curve. We finally separate into two regimes depending on whether or not they can collapse around the maximum diffusivity (dotted line), set by the gas turbulent $\alpha$ (see Eq. \eqref{eq:alpha}). We indicate dust sizes where collapse is possible regardless of diffusivity with the shaded regions. The area to the left of each shaded region but still below the diagonal curves indicates particle sizes where collapse remains possible, but only if the diffusivity is low enough.}
\label{fig:particleregime}
\end{figure*}

Another expectation from Equation \eqref{eq:gerbigparameter} is that the collapse of $\mathrm{St}=0.01$ dust appears to be difficult when the disk has a canonical metallicity of $Z=0.01$. In Figure \ref{fig:particleregime}, we explore some of the limitations on particle size $\mathrm{St}$ compared to the level of particle diffusion $\delta$. From Equation \eqref{eq:diffusionlimit}, we plot diagonal lines of constant $\epsilon Z$. These curves show that lowering the metallicity reduces the diffusivity needed to prevent gravitational collapse and the reverse is true when metallicity is increased. This is especially important when we consider the maximum diffusion possible, which is determined by the radial transport of grains and thus the $\alpha$ parameter for a self-gravitating disk with cooling timescale $\beta = 10$ \citep{Gammie2001}
\begin{equation}\label{eq:alpha}
\alpha_{\mathrm{GI}} = \frac{4}{9}\frac{1}{\gamma (\gamma - 1) \beta \Omega^{-1}} = 0.04.
\end{equation}
This divides the dust sizes into two regimes: those dust sizes that can collapse for any amount of turbulent diffusion one would reasonably expect in the disk (indicated by the different shaded regions and labeled as `ideal sizes' in Figure \ref{fig:particleregime}) and those that require that diffusion be lower or metallicity higher. At the same time, the choice of $\beta$ affects the strength of turbulence such that lower values of $\beta$ result in more efficient angular momentum transport which increases dust diffusion. This should push up the lower bound of `ideal sizes', limiting the size regime of particles that can freely collapse.

From this we surmise that metallicity should be a key constraint which determines the minimum grain size which can gravitationally collapse into a bound clump. For $Z=0.01$, this minimum is around $\mathrm{St} = 0.04$ and increases to $\mathrm{St}\sim 0.5$ for lower metallicities $Z=0.003$ while decreasing to $\mathrm{\mathrm{St}=0.005}$ for $Z=0.03$. Finally, it has been observed that a cut-off occurs at the higher end of dust sizes for $\mathrm{St} \gtrsim 1$ where dust never collapses \citep{Baehr2022,Rice2025}. While in \citet{Longarini2023b} and \citet{Rowther2024} simulations with larger $\mathrm{St}$ can form clumps, their models use grains of a constant size and the $\mathrm{St}$ number represents a simulation average and it may be the case that the grains closer to $\mathrm{St}\sim 1$ are forming clumps. While dust $\mathrm{St}\gtrsim1$ could collapse if it could concentrate enough, poor coupling and Equation \eqref{eq:stirringlimit} suggest that grains larger than 1 should no longer concentrate within spiral arms and instead be subject to gravitational interactions with the transient spiral density waves, leading to a gravitational stirring effect \citep{Shi2016}.

Streaming instability is another means of concentrating particles that can be compared to GI. Turbulence driven by SI is a few orders of magnitude weaker, so it is expected to form objects on the order of 100 km planetesimals (as in \citet{Li2021a}, \citet{Abod2019}, \citet{Gole2020}, etc.). GI-driven turbulence couples to dust particles most effectively at larger scales which leads to higher particle velocities and diffusion. As a result, the amount of mass that has to accumulate to overcome this diffusion for GI results in objects that are comparable to the Earth in mass. Furthermore, GI-driven turbulence is highly anisotropic \citep{Riols2020,Baehr2021a}, strongly favoring radial transport to vertical transport. If GI turbulence were more isotropic with efficient transport ($\alpha \sim 0.05$), the dust layer would be thicker, making it more difficult to reach critical densities necessary for clumping.

Disks are expected to be self-gravitating in the early stages of the disk lifetime. While dust growth can occur on short timescales, growth of grains larger than a few millimeters in the outer disk is not observed in significant quantities \citep{Ohashi2022}. If we assume a disk gas surface density distribution of $\Sigma_{\mathrm{g}} = 53 (R / 50\, \mathrm{au})^{-3/2}\, \mathrm{g\,cm}^{-2}$ we can convert our dimensionless particle size to a physical size
\begin{align}
a &= \frac{\mathrm{St}\, \Sigma_{\mathrm{g}}}{\sqrt{2\pi}\rho_{\bigcdot}}\\
&= 0.11 \left( \frac{\mathrm{St}}{0.01} \right) \left( \frac{R}{50\, \mathrm{au}} \right)^{-3/2} \mathrm{cm}.
\end{align}
Thus under our assumptions, dust sizes of $\mathrm{St} = 0.01$ correspond to approximately millimeter-sized grains in the outer regions of the disk, which may exist even in younger disks \citep{Zagaria2023,Doi2023}. Turbulent eddies in accretion disks are assumed to have turnover frequencies equal to the Keplerian frequency $\Omega$. However, it is possible that turbulent eddies have higher frequencies which would make smaller grains viable \citep{Sengupta2024}. There have been suggestions that dust growth can be rapid during the early stages of a protoplanetary disk, including during gravitational instabilities, producing large quantities of larger pebbles \citep{Vorobyov2023,Vorobyov2024}.

Simulations of a single grain size are valuable for keeping the dynamics of each size separate, making it easier to understand the nuances of each grain size. In real disks however, grains come in a wide range of sizes, typically dominated in number by micron sized grains \citep{Mathis1977}. Thus, a more realistic simulation would include particles over the entire size distribution and investigate how dust concentration operates with all the dust species active together, as was done in \citet{Rice2025}. When including a dust size distribution, they find that the formation of dense dust clouds is possible for typical disk metallicities ($Z=0.01$), which suggests that large grain sizes need only contribute a comparable amount of mass even if smaller grains are more numerous.

Although, as may be the case in a highly turbulent self-gravitating disk, large grains kicked to high velocities may lead to their own destruction via collisions \citep{Booth2016}, in which case it is still necessary to explore solids that are entirely made of small grains. The results of our simulations indicate that this limit is less problematic, as long as the dust content of the disk is increased.

While the formation of clumps is possible for small grains at $Z=0.04$, there is little evidence that metallicities are this high for an entire disk during the early stages of a protoplanetary disk. Measurements from evolved disks indicate that dust-to-gas ratios are sometimes higher than the ISM \citep{Miotello2017} and sometimes more lower than even this ISM value \citep{Trapman2025}, possibly indicating that dust has been processed into larger bodies such as planetesimals. Models of star and disk formation indicate that some regions may have moderately enhanced dust-to-gas ratios through redistribution of material ejected by the polar outflows \citep{Tsukamoto2021a}, natural settling of larger grains \citep{Lebreuilly2020a} or hydrodynamic instabilities during disk formation \citep{Bhandare2024}.

If direct formation of planetary cores in gravitoturbulent disks is a way of forming planets, then one would expect formation of wide orbit gas giants to be higher around stars with higher metallicities. Indeed, planet formation in general appears to be easier for stars with higher metallicity \citep{Petigura2018}, but whether this holds true for the population of cold gas giants is unclear \citep{Swastik2021}.

\section{Conclusion}
\label{sec:conclusion}

We have used hydrodynamic simulations with an operator splitting scheme to explore the behavior of small dust in self-gravitating disks. We find that small dust grains, on the order of a millimeter in size, are prone to gravitational collapse as long as the disk is enriched in dust above the typical ISM value of $Z=0.01$. 

We summarize our findings as follows:

\begin{itemize}
    \item For dust of sizes $\mathrm{St}=0.01$ to collapse into bound clumps, an increase of metallicity by a factor of at least 2 above solar is required. An initial metallicity of $Z=0.04$ produces more clumps of dust and on shorter timescales than $Z=0.02$. $\mathrm{St}=0.01$ corresponds to approximately 1$\mathrm{mm}$ grains at 50 au in a young disk and thus constitutes a more realistic grain size for a young disk.
    \item Even smaller grains with $\mathrm{St}=10^{-3}$ can also gravitationally collapse into a small number of bound clumps, but require $Z=0.04$ and even more time compared to $\mathrm{St}=0.01$ grains.
    \item Clump masses are typically between 0.1 and 1 $M_{\oplus}$, making them potential embryos or cores to planets in regions where disk self-gravity is relevant. Compared to simulations with larger grains, the clumps formed here have a higher minimum threshold, and between the two particle sizes used here, there is a factor of about two between the clump minimum mass.
    \item We find that the dust surface density enhancement $\Sigma_d/\langle \Sigma_d \rangle$ should be around 100 within the dense gas structures of a gravitationally unstable disk in order for the dust to collapse. This is independent of particle size or metallicity.
    \item Which dust grain sizes that can gravitationally collapse is determined by the strength of diffusion as a result of gas turbulence. For grains sizes larger than the dimensionless diffusion constant, the limit for which is the turbulent viscosity $\mathrm{St} > \delta \approx \alpha$, collapse is likely. Particles smaller than this threshold may only collapse under their own gravity with higher metallicities or a lower turbulent threshold.
\end{itemize}

\medskip

\begin{acknowledgments}
Simulations were conducted on the Vera supercomputer hosted by the Max Planck Computing and Data Facility (MPCDF).
CCY acknowledges the support from NASA via the Emerging Worlds program (\#80NSSC23K0653), the Astrophysics Theory Program (grant \#80NSSC24K0133), and the Theoretical and Computational Astrophysical Networks (grant \#80NSSC21K0497).
CH acknowledges support from NSF AAG grant No. 2407679 and from the National Geographic Society.
\end{acknowledgments}

\software{Matplotlib \citep{Hunter2007}, SciPy \& NumPy \citep{Virtanen2020,vanderWalt2011}, IPython \citep{Perez2007}}

\bibliography{library.bib}

\appendix

\section{Convergence}

\begin{figure}[h]
\centering
\includegraphics[width=0.48\textwidth]{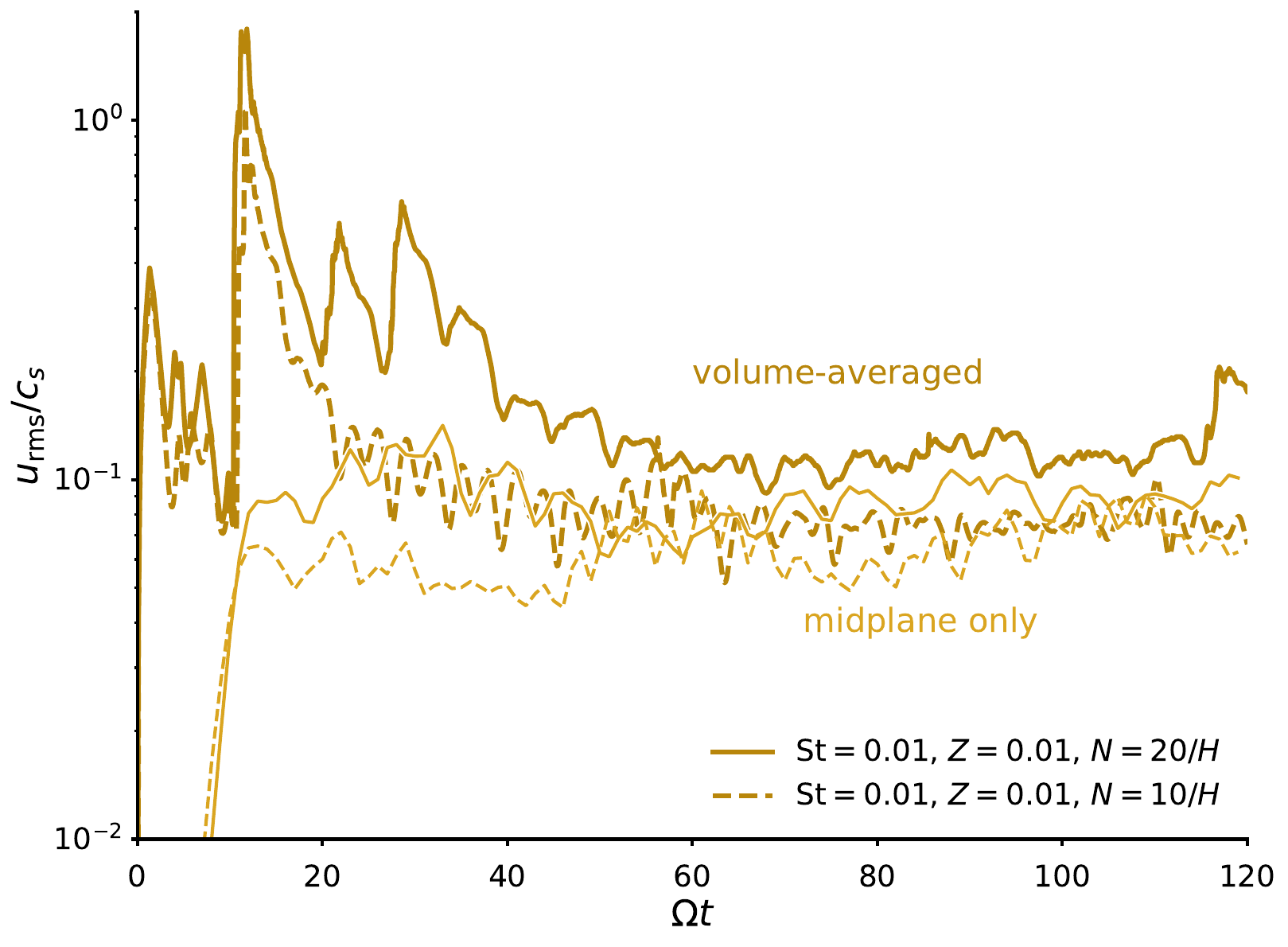}%
\includegraphics[width=0.48\textwidth]{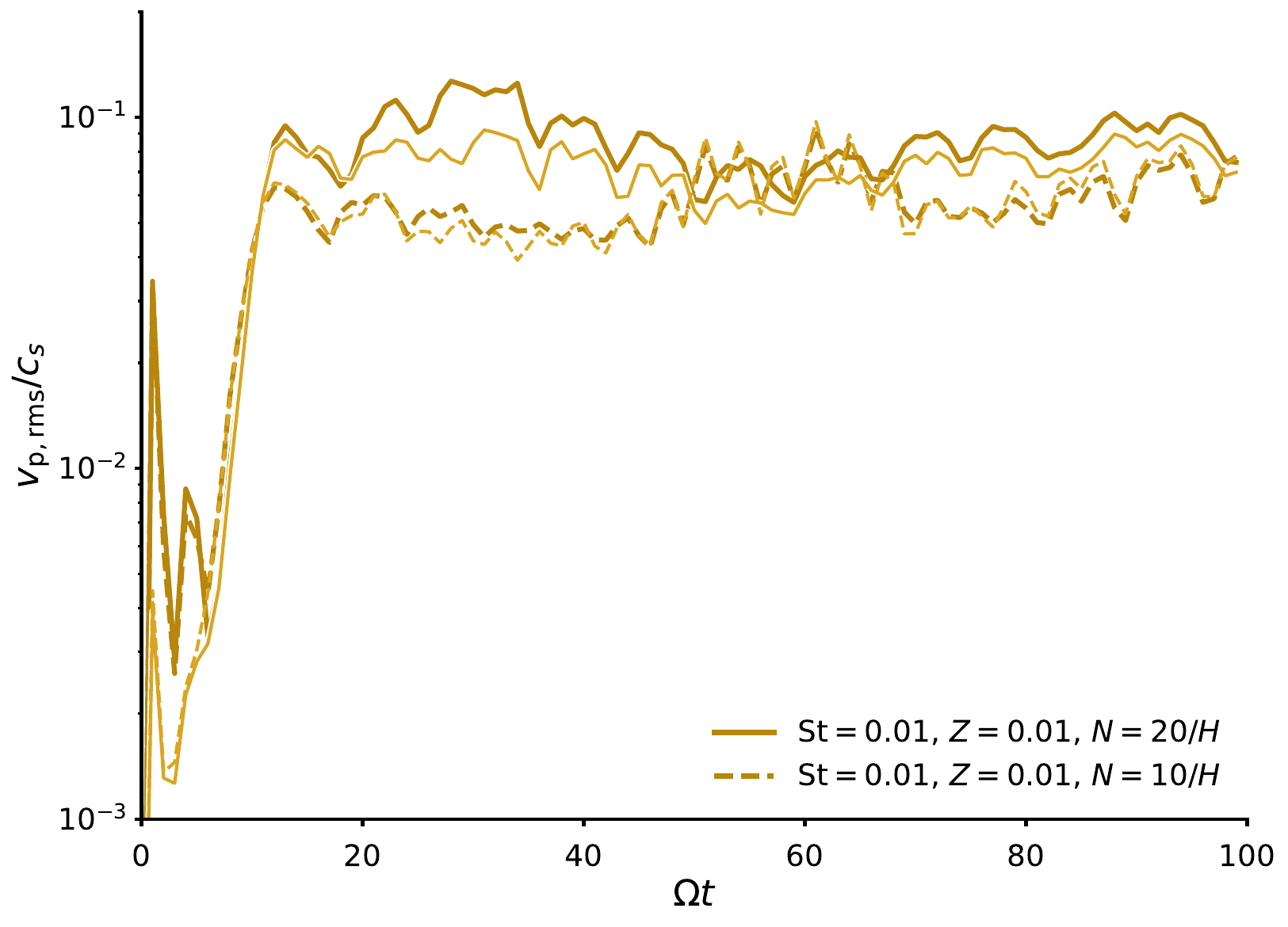}
\caption{\emph{Left:} Volume-averaged and midplane gas rms velocities in two simulations with different grid resolutions. The solid line shows the simulations with the nominal resolution $20/H$ while the dashed line shows the same for an identical simulation with $10/H$. Thinner and lighter lines show the rms velocity of the midplane layer only. \emph{Right:} Same as left panel but for dust rms velocities.}
\label{fig:rms}
\end{figure}

While \citet{Baehr2022} investigated whether higher resolutions had an effect on the mass distribution of clumps that formed and found that there was minimal impact (see Figure 10 therein), the convergence of the gas and dust rms velocities and kinetic energy spectra were not inspected. The main simulations of this paper were conducted at 20 cells per scale height. Double this resolution would have proven to be too time-consuming, so we opt instead to compare with lower resolution simulations at 10 cells per scale height. We limit the comparison here to the simulation which did not form any dense clumps of dust, but note that other runs were similar as long as clumps have not formed. In Figure \ref{fig:rms}, we show the rms velocities of both the gas and the dust for resolutions of 10 and 20 cells per scale height, either averaged over the whole domain or only over the midplane. It appears that each diagnostic has an intrinsic time variability of about a factor of two, and the differences between the two are also with about a factor of two. Therefore, or resolution of 20 cells per scale height might be close to the converged result.

\begin{figure}[t]
\centering
\includegraphics[width=0.48\textwidth]{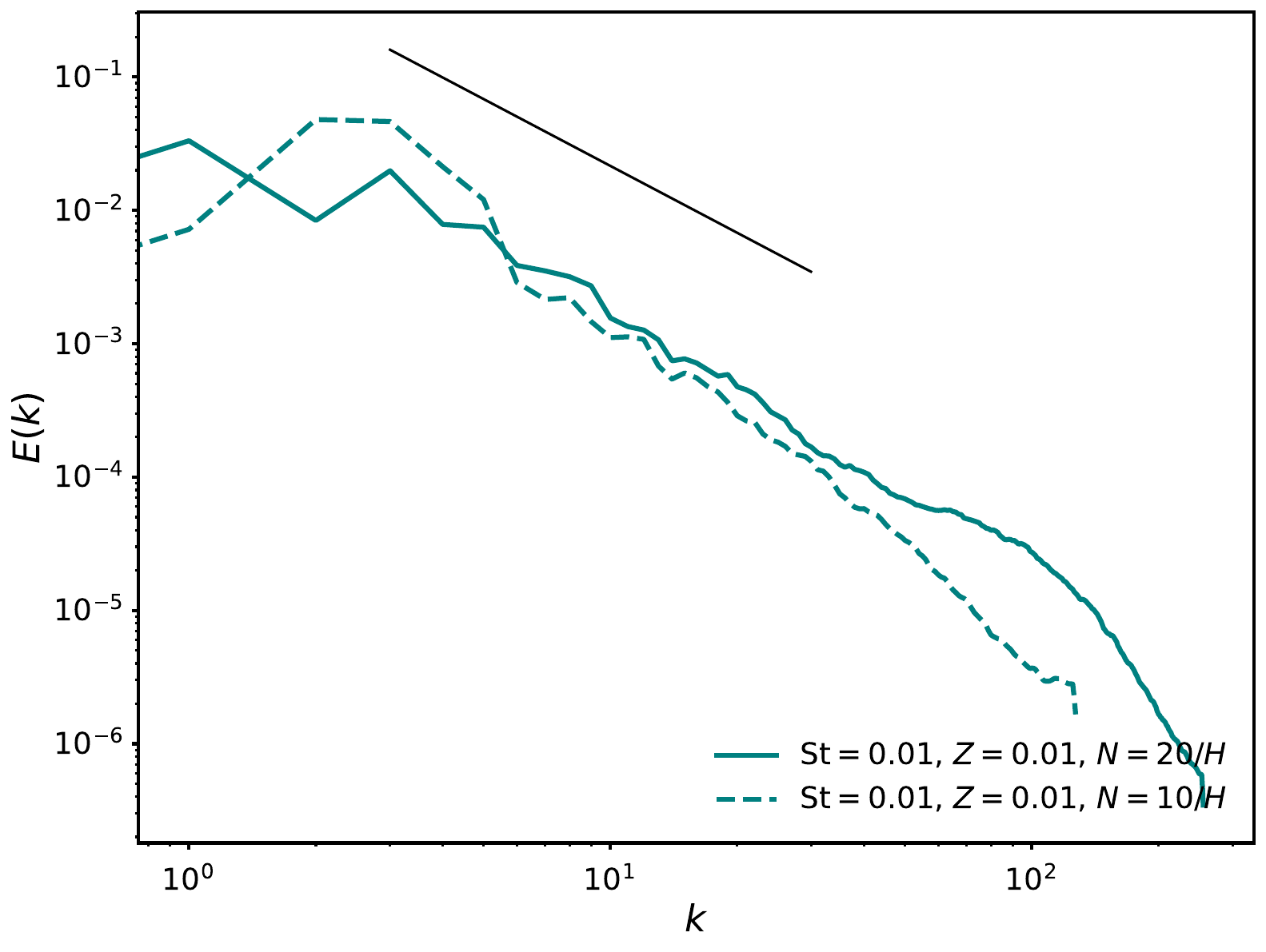}%
\includegraphics[width=0.48\textwidth]{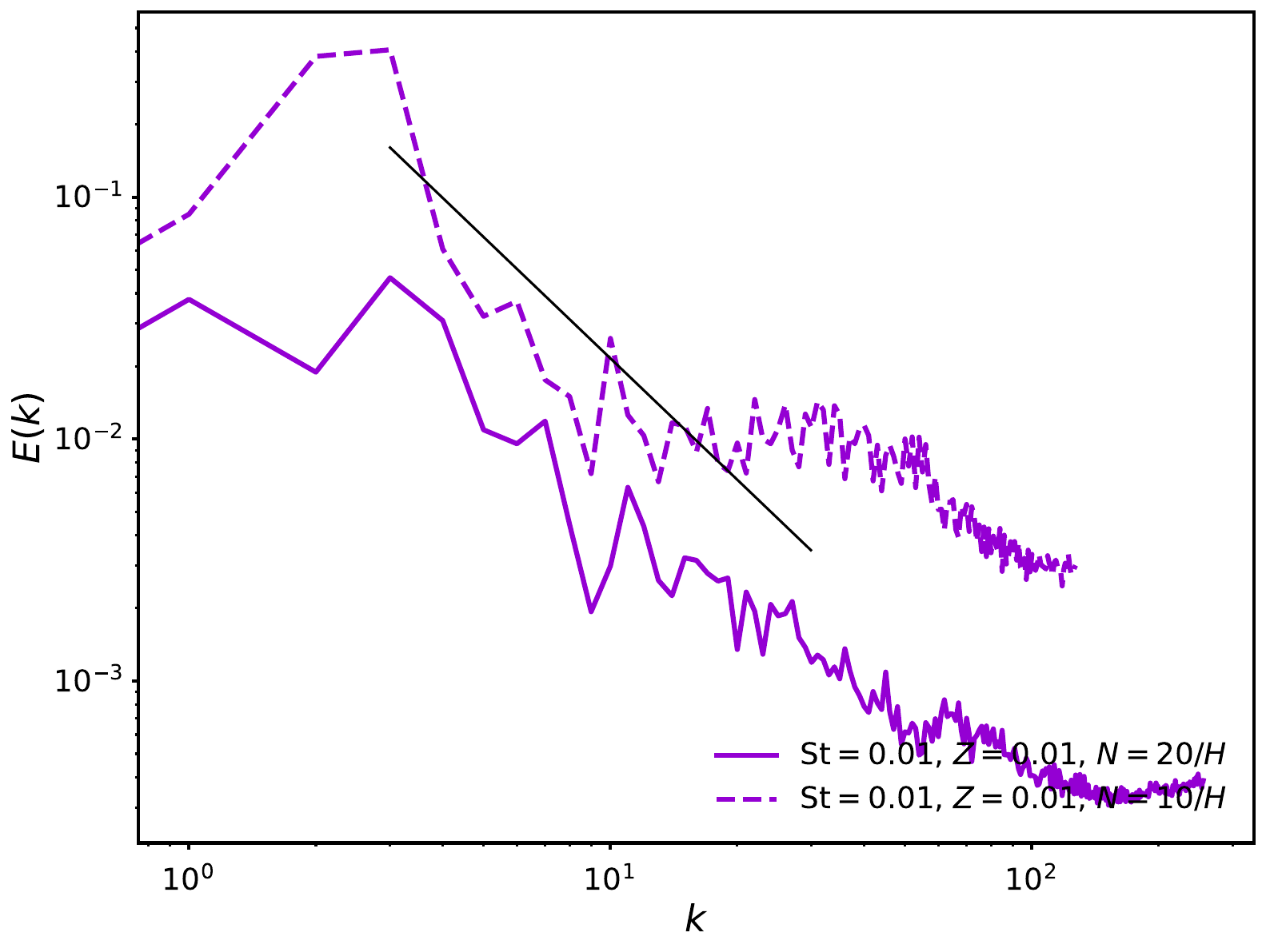}
\caption{\emph{Left:} Kinetic power spectrum of the gas around the midplane in the simulation small\_Z001 and its lower resolution counterpart, averaged over three snapshots at $t=50\Omega^{-1}$, $t=54\Omega^{-1}$ and $t=58\Omega^{-1}$. \emph{Right:} Similarly, the kinetic energy spectrum of the dust particles around the midplane. The thin black line indicates a slope of $k^{-5/3}$.}
\label{fig:spec}
\end{figure}

We calculate the 2D kinetic energy using the midplane velocities
\begin{equation} \label{eq:}
E(k) = \frac{1}{2}\tilde{u}\tilde{u}^{*},
\end{equation}
where $\tilde{u}$ is the Fourier transform $\mathcal{F}$ of velocity $u$
\begin{equation} \label{eq:fft}
\tilde{u}(k_x,k_y) = \mathcal{F}(u(x,y)) = \frac{1}{N_x N_y}\sum_{i=0}^{N_x - 1} \sum_{j=0}^{N_y - 1} u(x_i,y_j)e^{-ik_x x_i}e^{-ik_y y_j},
\end{equation}
and $^*$ denotes the complex conjugate and $k = (k_x^2 +k_y^2)^{1/2}$. In Figure \ref{fig:spec}, we show the resulting gas kinetic energy spectra, where the spectra are similar between resolutions of 10 and 20 cells per scale height. This is similar to \citet{Booth2019}, which also found convergence for the density-weighted gas kinetic energy spectra at the midplane. The simulation domain was chosen with the results of \citet{Booth2019} in mind, who showed that boxes that are too small in $x$ and $y$ may omit the larger unstable modes, while also keeping the domain small enough to modestly resolve the dust layer. Our choice of 25 $H$ captures around 4 times the most unstable wavelength, $\lambda \sim 2\pi H$, while not fully including 60$H$ may result in some bursty behavior. Using the full vertical extent shows that both spectra have unphysical signatures due to the incomplete vertical domain. This reflects the importance of the dense midplane with regards to gravitoturbulence \citep[as in][]{Shi2014}. The dust particles are however not converged, and have a shallower cascade on smaller scales.

The reason why the kinetic energy spectra of the dust have not reached convergence remains to be seen, as previous studies have shown that the cascade of the particle kinetic energy spectrum can vary with resolution and dust properties \citep{Sengupta2023}. The dust layer becomes increasingly thin as the dust settles to the midplane and perhaps no longer adequately resolved by the grid resolution. This suggests that while the gas dynamics of gravitoturbulence is reasonably well-resolved, the onset of dense region of dust may not reliably model the dust dynamics. Higher resolution is needed of the particle layer in particular, i.e. both more grid cells to cover the midplane as well as more particles within the layer.

\end{document}